\documentclass[aps,nofootinbib,prd,eqsecnum,showpacs,showkeys,preprintnumbers]{revtex4-2}
\usepackage{eurosym}
\usepackage[caption=false]{subfig}
\usepackage{graphicx}
\usepackage{amsmath}
\usepackage{amsfonts}
\usepackage{amssymb}
\usepackage{float}
\usepackage{color}
\usepackage{bm}
\usepackage{mathrsfs}
\usepackage{epstopdf}
\usepackage{url}
\usepackage{footnote}
\usepackage{textcomp}
\usepackage{ulem}
\usepackage{esint}
\usepackage[unicode=true, pdfusetitle,
 bookmarks=true,bookmarksnumbered=false,bookmarksopen=false,
 breaklinks=false,pdfborder={0 0 1},backref=false,colorlinks=false]{hyperref}
\usepackage{multirow}
\usepackage{pifont}

\begin{document}

\title{Tracing Inflationary Imprints Through the Dark Ages: Implications for
Early Stars and Galaxies Formation}
\author{K. El Bourakadi$^{1,2}$}
\email{k.elbourakadi@yahoo.com}
\author{M. Yu. Khlopov$^{3}$}
\email{khlopov@apc.in2p3.fr}
\author{M. Krasnov$^{4,5}$}
\author{H. Chakir$^{2}$}
\email{chakir10@gmail.com }
\author{M. Bennai$^{1}$}
\email{mdbennai@yahoo.fr}
\date{\today }

\begin{abstract}
We explore how inflationary features shape the early stages of cosmic
structure formation. Using the transfer function formalism, we trace the
evolution of primordial perturbations, showing how causal physics and
oscillatory signatures from inflation influence the matter power spectrum.
The variance of smoothed density fields is then applied to model the
collapse of overdense regions and predict dark matter halo abundances
through the Press--Schechter framework. Extending to the baryonic sector, we
analyze primordial gas collapse in minihalos, emphasizing molecular hydrogen
cooling and the thermochemical pathways leading to Population III star
formation. Finally, we examine primordial black holes as potential seeds for
early galaxies, connecting their accretion-driven growth to the stellar
masses and disk properties of high-redshift systems. Our results indicate
that oscillatory features from inflation can leave measurable imprints on
halo abundances and early galaxy properties, providing a testable link
between high-energy physics and astrophysical observations with JWST.
\end{abstract}

\affiliation{$^{1}$\small Physics and Quantum Technology Team, LPMC, Ben
M’sik Faculty of Sciences, Casablanca Hassan II University, Casablanca,
Morocco} 
\affiliation{$^{2}${\small Subatomic Research and Applications Team, Faculty of Science
Ben M'sik,}\\
{\small Casablanca Hassan II University, Morocco}\\} 
\affiliation{$^{3}$
Virtual Institute of Astroparticle Physics, 75018 Paris, France\\} 
\affiliation{$^{4}$
Institute of Physics, Southern Federal University, 194 Stachki, Rostov-on-Donu, Russia\\}
\affiliation{$^{5}$ 
National Research Nuclear University MEPhI, 115409 Moscow, Russia\\}

\maketitle

\section{Introduction}

%%%%%%%%%%%%%%%%%%%%%%%%%%%%%%%%%%%%%%%%%%%%%%

The study of cosmological perturbations has long been a cornerstone for
understanding the emergence of structure in the universe. Small primordial
inhomogeneities generated during inflation provide the seeds for galaxies
and large-scale structure, with their subsequent growth shaped by the
physical properties of matter and radiation across cosmic epochs \cite%
{Dodelson 2003,Mukhanov 2005}. The transfer function formalism captures how
these primordial fluctuations evolve under the combined effects of
causality, radiation pressure, and baryonic physics, acting as a
scale-dependent filter that links early curvature perturbations to
present-day matter distributions \cite{Bardeen 1987, Eisenstein 1998}.
During the radiation-dominated era, perturbations inside the horizon are
suppressed, while after matter-radiation equality, the growth of cold dark
matter overdensities dominates the gravitational potential and drives
structure formation \cite{Peacock 1999}. More recently, inflationary models
embedded in high-energy frameworks such as string theory, particularly
axion-monodromy scenarios, have been shown to produce distinctive
oscillatory signatures in the primordial power spectrum, potentially
detectable in cosmic microwave background and large-scale structure surveys 
\cite{McAllister 2014}. These combined developments emphasize the interplay
between fundamental physics at early times and the observable features
imprinted on the matter power spectrum today, motivating the detailed
investigation undertaken in this work.

The gravitational collapse of overdense regions represents the pivotal step
in the transition from linear cosmological perturbations to the nonlinear
structures observed in the universe today. Tiny primordial fluctuations,
imprinted during inflation and later detected in the cosmic microwave
background, provided the seeds for galaxies, clusters, and dark matter halos 
\cite{Guth 1981,Starobinsky 1982,Planck 2018}. Their subsequent growth
through gravitational instability is well described by linear perturbation
theory in the early stages \cite{Peebles 1982, Dodelson 2003}, but nonlinear
effects eventually dominate, necessitating more advanced modeling. The fluid
approximation captures the large-scale behavior of both dark matter and
baryons \cite{Bertschinger 1995, Mo 2010}, while spherical collapse models
provide an analytically tractable framework for studying halo formation \cite%
{Gunn 1972, Padmanabhan 1993}. Despite their simplicity, these models
reproduce many of the key properties seen in simulations and observations,
including the abundance and clustering of halos \cite{Press 1974, Bond 1991,
Springel 2005}. Central to this framework is the concept of a critical
linear overdensity threshold $\delta _{c}\approx 1.686$, which marks the
point at which regions collapse under their own gravity \cite{Peebles 1987,
Eke 1996}. This criterion, together with the mass variance $\sigma \left(
M\right) $, underpins predictions of the halo mass function, linking
primordial inflationary fluctuations to the statistical distribution of
collapsed structures \cite{Barkana 2001, Wechsler 2018, El Bourakadi 2023,
El Bourakadi 2022-1, El Bourakadi 2021-1, Bousder 2021, El Bourakadi 2022-2,
El Bourakadi 2021-0, Sakhi 2020, El Bourakadi 2024-1, El Bourakadi 2024-2,
Hanin 2023,  El Bourakadi 2024-3, El Bourakadi 2026}. Thus, understanding
the collapse of overdense regions not only explains the emergence of dark
matter halos but also provides a bridge between early-universe physics and
the nonlinear growth of cosmic structure.

The formation of the first stars, often referred to as Population III, marks
a fundamental turning point in cosmic history, transforming the universe
from a dark, simple state into one enriched by radiation and heavy elements.
Within the $\Lambda $CDM framework, calibrated by high-precision
observations of the cosmic microwave background \cite{Larson 2011, Planck
2018, Planck 2020}, these stars are predicted to form in dark matter
minihalos of mass $\sim 10^{5}-10^{6}M_{\odot }$ at redshifts $z\sim 20-30$.
The collapse of pristine baryonic gas into such halos was initially modeled
by spherical collapse theory \cite{Gunn 1972, Tegmark 1997}, which provides
analytic estimates of virial radii and temperatures \cite{Barkana 2001}. The
cooling of primordial gas was governed primarily by molecular hydrogen,
forming through non-equilibrium processes in the absence of dust grains \cite%
{Galli 1998, Hollenbach 1979}, and regulating the thermodynamics of collapse
to a few hundred kelvin. Simulations have since revealed that the chemistry
and dynamics of primordial gas were more intricate than initially assumed,
progressing through multiple density-dependent phases that ultimately led to
the runaway collapse of pre-stellar cores \cite{Abel 2002, Bromm 2004,
Yoshida 2006}. At higher densities, three-body reactions drove rapid $H_{2} $
formation \cite{Palla 1983}, while radiative transfer effects required
sophisticated approaches such as the Sobolev approximation to capture
cooling accurately \cite{Omukai 1998, Greif 2012}. At the end of this
sequence, a protostellar core of $\sim 100-200$ $R_{\odot }$ emerged \cite%
{Stahler 1986, Stacy 2013}, initiating the first episode of stellar
evolution and feedback in the universe. The study of these processes not
only links early-universe physics with baryonic collapse,, but also
establishes the initial conditions for subsequent galaxy formation and
cosmic reionization.\newline

Inflationary features that modulate small-scale power can leave distinctive,
testable imprints on both the abundance of primordial black holes (PBHs) and
the properties of nascent galaxies. In scenarios where the inflaton dynamics
generate enhanced curvature fluctuations---or produce metastable domains
bounded by collapsing vacuum walls---PBHs may form in the early universe and
later act as gravitational seeds for structure growth \cite{Khlopov 2002,
Khlopov 2005, Khlopov 2007, Barbieri 2024}. Their cosmological impact is
twofold: as a population, they source isocurvature-like \textquotedblleft
Poisson \textquotedblright\ fluctuations and local \textquotedblleft seed
\textquotedblright\ over-densities that accelerate halo assembly, while
individually they can grow by gas accretion inside forming halos \cite{Hoyle
1966, Carr 1983, Carr 1984, Carr 2018, De Luca 2020, De Luca 2020, De Luca
2023, Yuan 2024}. Observational bounds from the CMB, Lyman-$\alpha $ forest,
X-ray binaries, Galactic dynamics and large-scale structure restrict the
allowed PBH mass--fraction across wide mass ranges, yet still permit
subdominant PBH populations capable of influencing early galaxy formation 
\cite{Sanchez 1997, Green 1997, Murgia 2019, Su 2023, Carr 2018}.
Intriguingly, JWST detections of luminous galaxies at $z\gtrsim 10$ and
candidate early active nuclei motivate renewed consideration of massive
seeds and rapid early growth channels \cite{Labbe 2023, Volonteri 2012,
Curti 2023, Bouwens 2010, Maiolino 2024, Terao 2022}. In parallel, the
macroscopic properties of young galaxies --stellar mass, size, and disk
scale radius---can be linked to their host halos through simple
angular-momentum--conserving models, providing a clean conduit from initial
conditions to observable structure \cite{Fall 1980, Dalcanton 1997, Mo 1998}%
. Building on these ideas, this section quantifies how inflation-induced
oscillations in the primordial spectrum regulate PBH seeding and accretion
(via Eddington-limited growth \cite{Eddington 2013, Valentini 2020}) and,
through their impact on halo assembly, propagate into the stellar masses and
disk sizes of early galaxies. \newline

This work presents an analysis of the impact of inflationary features
arising from axion-monodromy inflation on cosmic structure formation during
the Dark Ages. Axion-monodromy inflation is a well-established
string-inspired framework that naturally generates oscillatory signatures
and localized enhancements in the primordial power spectrum \cite%
{Silverstein 2008, Flauger 2010, Easther 2014, Flauger 2017}. Previous
investigations have primarily focused on the phenomenology of these
features, particularly their role in primordial black hole formation and the
resulting constraints on PBH abundance driven by amplified small-scale power 
\cite{Inomata 2017, Inomata 2021, Ballesteros 2018, Ozsoy 2018}. Such
studies, however, typically address isolated components of the problem such
as the shape and amplitude of the primordial power spectrum or PBH abundance
without explicitly tracking the subsequent nonlinear evolution of matter
perturbations into dark matter halos and the associated astrophysical
processes. Although modifications of the primordial power spectrum and their
effects on small-scale halo populations have been examined independently 
\cite{Gilman 2021}, a direct connection to concrete inflationary models
remains largely unexplored. The novelty of the present work lies in bridging
this gap by consistently following axion-monodromy--induced features from
the primordial power spectrum through nonlinear halo collapse, star
formation, PBH formation, and early galaxy assembly, thereby establishing a
coherent link between inflationary physics and observable signatures of
structure formation in the early Universe. \newline

In what follows, we build on these foundations by examining in detail the
connections between early-universe physics and the emergence of cosmic
structure. In Section \ref{sec2} we develop the formalism of cosmological
perturbations through the transfer function. In Section \ref{sec3} we
address the gravitational collapse of overdense regions, establishing the
statistical framework for dark matter halo formation. In Section \ref{sec4}
we extend this analysis to the baryonic sector, tracing the collapse of
primordial gas within minihalos, the thermochemical regulation by molecular
hydrogen, and the onset of Population III star formation. Finally, in
Section \ref{sec5} role of primordial black holes as seeds for early
galaxies, exploring their growth via accretion, their impact on
high-redshift galaxy properties, and the imprint of inflationary features on
galactic disk formation. We conclude in Section \ref{sec6}.

\section{Evolution of Cosmological Perturbations}

\label{sec2}

\subsection{The Transfer Function}

The evolution of cosmological perturbations is strongly influenced by the
physical properties of the universe's constituents, which can significantly
modify predictions from perturbation theory, leading to deviations from
idealized models \cite{Dodelson 2003,Mukhanov 2005}. However, causality
imposes fundamental constraints, preventing superluminal propagation of
influence beyond the cosmological horizon \cite{Ellis 1993}. To quantify
these effects, standard practice compares observed perturbations with those
predicted in scenarios where causal physics is neglected \cite{Peebles 1987}%
. The resulting power spectrum, scaling with the square of the transfer
function $T(k)$ multiplied by the primordial power spectrum \cite{Bardeen
1987}. This relationship underscores the role of microphysical processes in
large-scale structure formation \cite{Weinberg 2008}. The transition from a
radiation-dominated universe to a matter-dominated one occurs roughly at 
\cite{Eisenstein 1998},

\begin{eqnarray}
z_{eq} &=&2.50\times 10^{4}\Omega _{m}h^{2}\Theta _{2.7}^{-4} \\
k_{eq} &\equiv &\left( 2\Omega _{m}H_{0}^{2}z_{eq}\right) ^{1/2}
\end{eqnarray}%
where $z_{eq}$ and\ $k_{eq}$\ are the redshift and the scale at the
transition step, $\Theta _{2.7}^{-4}$\ refers to the dimensionless CMB
temperature fluctuation, $\Omega _{m}$\ is the total matter density
containing the baryons and dark matter, and $h$\ is the dimensionless Hubble
constant. Analytic solutions exist for linear perturbations at small scales
when baryons' gravitational influence on CDM is negligible \cite{Hu 1996}.
This holds below the sound horizon, where baryon pressure dominates \cite%
{Eisenstein 1998}. The transfer function at the drag epoch \cite{Ma 1995},
forming our small-scale fitting basis \cite{Eisenstein 1998, Dodelson 2024},%
\begin{eqnarray}
T\left( k\right) &=&\frac{15}{4}\frac{\Omega _{m}H_{0}^{2}}{k^{2}a_{eq}}A\ln %
\left[ \frac{4Be^{-3}\sqrt{2}k}{k_{eq}}\right] , \\
&=&12\frac{k_{eq}^{2}}{k^{2}}\ln \left[ 0.12\frac{k}{k_{eq}}\right] .
\end{eqnarray}

The analytic transfer function approximation provides a valid framework for
studying gravitational collapse during star formation when applied to
small-scale perturbations $\left( k\gtrsim 0.1h~Mpc^{-1}\right) $\ in the
post-recombination era. While originally derived for linear cosmological
perturbations \cite{Eisenstein 1998}, this formalism remains applicable to
local collapse scenarios because: (1) it captures the critical Jeans-scale
suppression of gas perturbations due to pressure support \cite{Jeans 1902},
(2) the matter-dominated growth factor $D(z)$ correctly describes the
gravitational amplification of overdensities prior to nonlinear collapse 
\cite{Peebles 1982}, and (3) the separation of scales allows baryonic
physics to be incorporated through modified initial conditions \cite{Schaye
2008}. For protostellar collapse $\left( 10^{2}-10^{5}M_{\odot }\right) $
scales, the transfer function's $k$-dependence replicates how gas pressure
filters out small fluctuations while permitting larger-scale collapse \cite%
{Tseliakhovich 2010a, Tseliakhovich 2010b}. This is particularly evident
when comparing the characteristic turnover in $T\left( k\right) $ to the
observed mass function of Population $III$ stars \cite{Hirano 2014}.
However, the approximation requires augmentation with: (a) radiative cooling
terms for $z\lesssim 100$, and (b) nonlinear corrections when $\delta >1$ 
\cite{Press 1974}, making it most accurate for studying the initial phases
of gravitational instability before fragmentation dominates.

\subsection{The Matter Perturbations}

During the radiation-dominated epoch, horizon entry causes gravitational
potentials to decay due to radiation pressure effects \cite{Dodelson 2024},
which strongly suppress growth of radiation density perturbations \cite%
{Mukhanov 2005}. Initially, radiation dominates both the background energy
density and the gravitational potential \cite{Peacock 1999}. However, the
continuous growth of matter perturbations eventually leads to a crucial
crossover despite matter's lower background density \cite{Ma 1995}. Beyond
this point, the gravitational potential becomes governed primarily by the
dark matter distribution, and their subsequent evolution becomes decoupled
from radiation dynamics \cite{Hu 1996}. This transition represents a
critical phase in structure formation, marking the shift from
radiation-dominated to matter-dominated perturbation growth in the early
universe \cite{Planck 2018}.

The transfer function $T(k)$\ encodes how primordial curvature perturbations 
$\mathcal{R}_{k}$ (generated during inflation \cite{Planck 2018}) evolve
into matter density perturbations $\delta _{c}$ during formation of cosmic
structure. As demonstrated by \cite{Dodelson 2024}, in the matter-dominated
era $\left( a\gg a_{eq}\right) ,$\ the growing mode of $\delta _{c}$\
dominates, while the decaying mode becomes negligible. Following the
formalism of \cite{Eisenstein 1998}, we can assemble these results to derive
the transfer function's analytical form. Specifically, on small scales, the
matter density perturbation \ is directly linked to the initial curvature
perturbation through the transfer function \cite{Ma 1995}, with the transfer
function accounting for suppression effects from: $\left( 1\right) $\
radiation domination (via the Meszaros effect \cite{Mukhanov 2005}), and $%
(2) $ baryonic physics \cite{Barkana 2001}. In the limit $a\gg a_{eq},$ $%
\delta _{c}$ simplifies to a form where $T(k)$ acts as a scale-dependent
filter converting primordial fluctuations to late-time structure \cite%
{Weinberg 2008}, as confirmed by numerical solutions of the
Einstein-Boltzmann equations \cite{Blas 2011}. Therefore, at late times the
solution for the small-scale perturbations is approximated \cite{Dodelson
2024},

\begin{equation}
\delta _{c}(k,z)=\frac{2}{5}\frac{k^{2}}{\Omega _{m}H_{0}^{2}}\mathcal{R}%
(k)T\left( k\right) D(z),
\end{equation}%
knowing that $a=\left( 1+z\right) ^{-1},$\ $\left\langle \left\vert \delta
_{c}(k,z)\right\vert ^{2}\right\rangle =\mathcal{P}(k,z),$and \ $\Delta _{%
\mathcal{R}}^{2}(k)=\frac{k^{3}}{2\pi ^{2}}\mathcal{R}\left( k\right) .$

\subsection{Axion-Monodromy Inflation with Oscillatory Features}

Axion-monodromy inflation provides a theoretically well-motivated framework
for embedding inflationary dynamics within string theory. This class of
models is characterized by an inflaton potential of the form \cite%
{Silverstein 2008, McAllister 2010}%
\begin{equation}
V\left( \phi \right) =V_{0}\phi +\Lambda ^{4}\cos \left( \frac{\phi }{f}%
\right) ,
\end{equation}%
where the power-law term drives slow-roll inflation, while the superimposed
cosine encodes periodic modulations resulting from monodromic effects in
axionic fields, this potential arises from the breaking of the axion shift
symmetry in string compactification, leading to observable imprints on the
primordial power spectrum. Specifically, the oscillatory term generates
scale-dependent modulations in the scalar perturbations, which may manifest
as resonant or sharp features in the cosmic microwave background, as well as
in large-scale structure tracers such as the Lyman-$\alpha $ forest and
21-cm intensity mapping \cite{Munoz 2020,Chen 2016}. These features have
been analyzed and constrained in works such as \cite{Planck 2018, Flauger
2010, Stahl 2025, Liu 2021}, and are implemented in numerical tools like
CLASS and N-body simulations for forecasting and data comparison \cite%
{Lesgourgues 2014}. Writing the equation of motion one can find,%
\begin{equation}
\ddot{\phi}+3H\dot{\phi}+V_{0}-V_{0}b\sin \left( \frac{\phi }{f}\right) =0,
\end{equation}%
with $b\equiv \frac{\Lambda ^{4}}{V_{0}f}.$\ The solution to the Klien
Gordan Equation is found to be \cite{Flauger 2010},%
\begin{equation}
\phi (t)=\phi _{0}(t)+b\phi _{1}(t),
\end{equation}%
here%
\begin{eqnarray}
\phi _{0}(t) &=&\left[ \phi _{\ast }^{3/2}-\frac{\sqrt{3}}{2}%
V_{0}^{1/2}\left( t-t_{\ast }\right) \right] ^{2/3}, \\
\phi _{1}(t) &=&f\frac{3f\phi _{\ast }}{1+\left( 3f\phi _{\ast }\right) ^{2}}%
\left[ -\sin \left( \frac{\phi _{0}(t)}{f}\right) +3f\phi _{\ast }\cos
\left( \frac{\phi _{0}(t)}{f}\right) \right] .
\end{eqnarray}%
Here, $\phi _{\ast }$\ denotes the value of the field $\phi _{0}$\ at the
time at which the pivot scale $k_{\ast }$\ exits the horizon. In the absence
of oscillations $(b=0)$, axion monodromy inflation behaves as a standard
large-field model, well-described by slow-roll approximations \cite%
{Silverstein 2008}. For CMB scales exiting the horizon $60$ e-folds before
inflation ends. When oscillations are introduced $(b\neq 0)$, while the
background evolution and total e-folds remain largely unaffected for small
b, the perturbation spectrum develops significant non-slow-roll features
requiring specialized analysis beyond the slow-roll approximation.

Having established the background evolution, the power spectrum was computed
analytically under the same assumptions: slow-roll for $\phi _{0}(t) $, $%
\phi _{0}(t)\gg M_{p}$, $f\ll M_{p}$, and to the first order in $b$. The
scalar power spectrum takes the form:%
\begin{equation}
\Delta _{\mathcal{R}}^{2}(k)=\Delta _{\mathcal{R}}^{2}(k_{\ast })\left( 
\frac{k}{k_{\ast }}\right) ^{n_{s}-1}\left[ 1+\delta n_{s}\cos \left( \frac{%
\phi _{k}}{f}\right) \right]
\end{equation}%
given that,%
\begin{equation}
\delta n_{s}=\frac{12b}{\sqrt{\left( 1+\left( 3f\phi _{\ast }\right)
^{2}\right) }}\sqrt{\frac{\pi }{8}\coth \left( \frac{\pi }{2f\phi _{\ast }}%
\right) f\phi _{\ast }},
\end{equation}%
where $\phi _{k}$\ is the value of the scalar field at the time when the
mode with comoving momentum $k$ exits the horizon. In the analysis by \cite%
{Flauger 2010}, the choice of the axion decay constant $f=0.01$ emerges as a
particularly instructive value to probe oscillatory features in the CMB
power spectrum arising from axion monodromy inflation. This value is
highlighted, since it lies within the range where the degeneracy between the
baryon density $\Omega _{B}h^{2}$ and the modulation amplitude $\delta n_{s}$%
{} becomes pronounced, as demonstrated in their work. This degeneracy
complicates parameter estimation but also provides a clear signature of
resonant non-Gaussianity, making $f=0.01$ a critical benchmark for
distinguishing axion monodromy from standard $\Lambda $CDM predictions. 
\begin{figure}[h]
\centering
\includegraphics[width=0.99\textwidth]{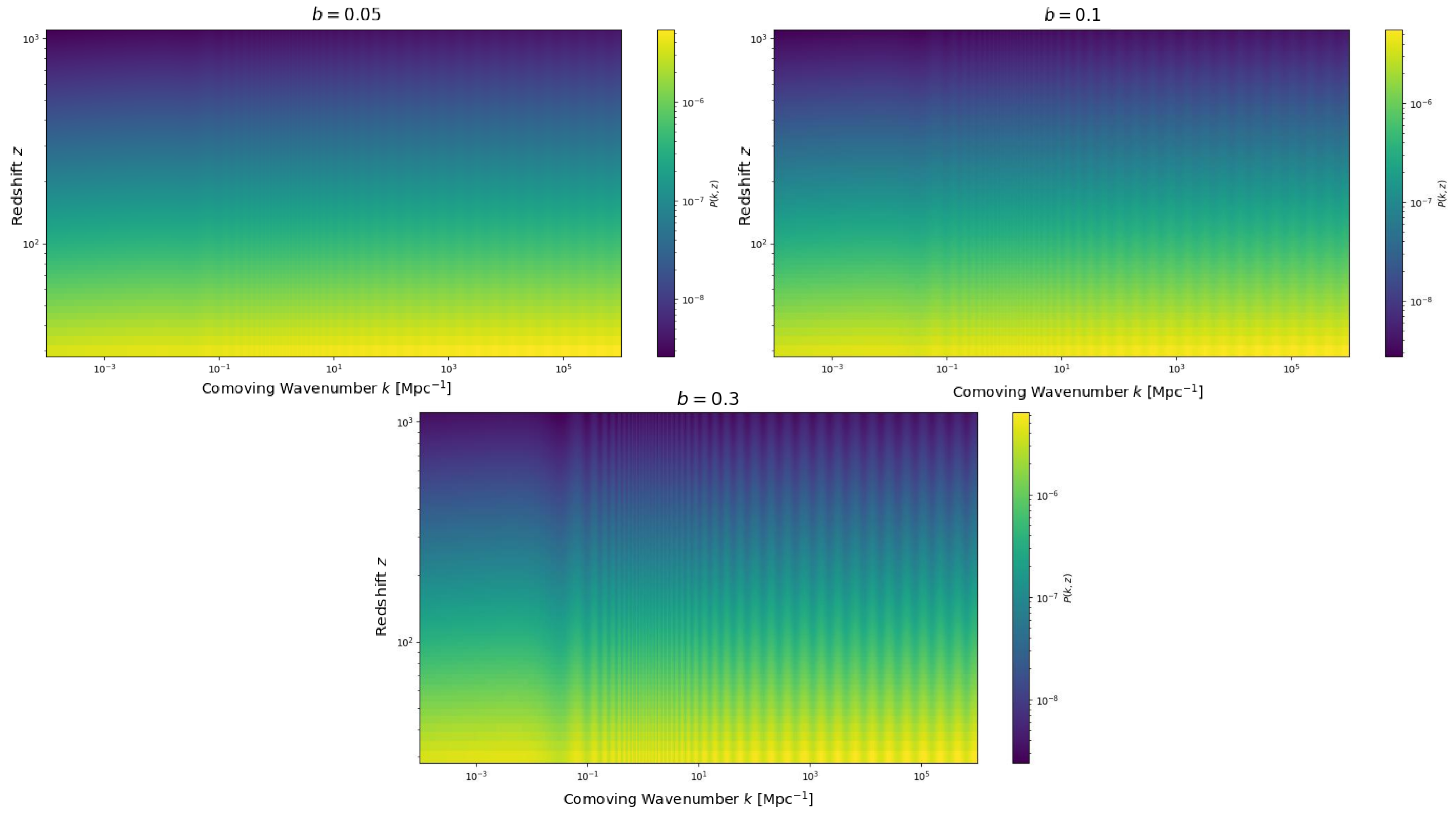} 
% Adjust width as needed
% Label for referencing the figure
\caption{ Matter power spectrum $P(k,z)$ plotted over comoving scale and
redshift for different values of b showing enhanced primordial oscillations }
\label{fig:1}
\end{figure}

The Fig. \ref{fig:1} visualizes evolution of the matter power spectrum $%
P(k,z)$ as a function of comoving wavenumber $k$ and redshift $z$. It
highlights variation of parameter $b$, which governs the amplitude of
primordial oscillations, modulating structure formation. All models include
an oscillatory correction in the primordial power spectrum stemming from
inflationary physics, with the oscillation strength increasing from $b=0.05$
to $b=0.3$. This leads to increasingly visible ripples in $P(k,z)$ for
certain intervals of $k$. These oscillations can be interpreted as imprints
of periodic features in the inflaton potential, such as those arising in
axion-monodromy models, and they propagate forward in time into the
large-scale structure distribution. Additionally, the redshift evolution
from the post-recombination epoch ($z\sim 1100$) through the dark ages ($%
z\sim 30$) reveals a steady suppression of the power spectrum amplitude.
This is driven by the growth factor $D(z)\propto (1+z)^{-1}$ \cite{Dodelson
2003, Eisenstein 1998}, which reflects the fact that matter perturbations
grow more efficiently at lower redshift. As a result, while the power
spectrum at high $z$ shows the primordial shape modulated by the transfer
functions and the oscillations, the late-time power spectrum becomes
increasingly amplified at low $k$, highlighting the gravitational clumping
of matter into cosmic structures. The impact of increasing $b$ becomes most
prominent at intermediate-to-small scales, suggesting that precise
measurements of power spectrum from galaxy surveys could be used to place
strong constraints on the amplitude and scale of primordial features.

\begin{figure}[h]
\centering
\includegraphics[width=0.99\textwidth]{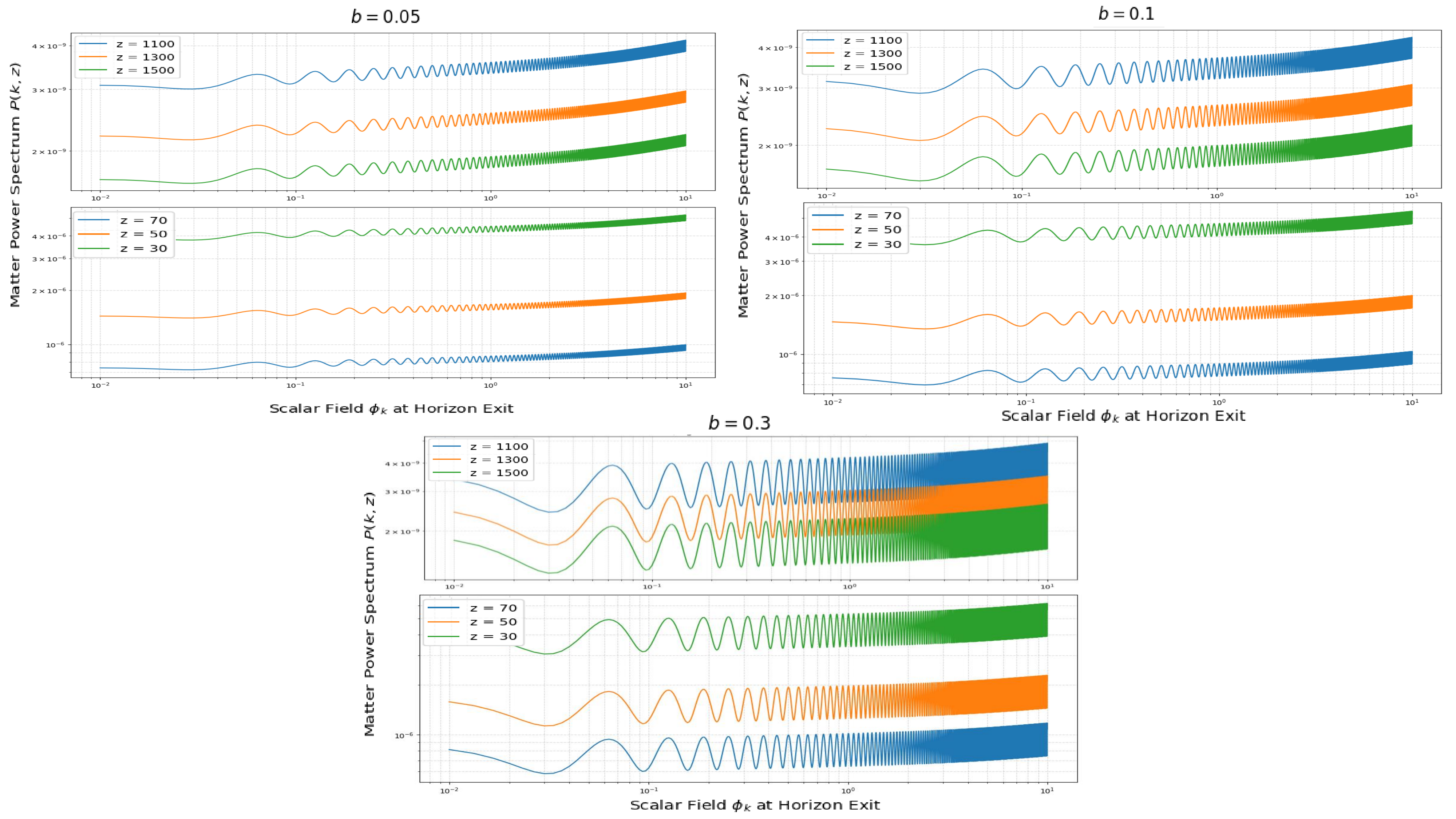} 
% Adjust width as needed
% Label for referencing the figure
\caption{ Matter Power Spectrum Evolution with Inflaton-Induced Oscillations
for different choices of Redshift values.}
\label{fig:2}
\end{figure}
The results presented in Fig.\ref{fig:2} explore the evolution of the matter
power spectrum $P(k,z)$ as a function of the inflaton field value $\phi _{k}$
at horizon crossing, for various redshifts spanning from the epoch of
recombination to the end of the dark ages. The parameter $b$ modulates the
amplitude of oscillations, superimposed on the primordial curvature
perturbation spectrum via $\delta n_{s}$. Physically, these oscillations
reflect possible periodic features in the inflationary potential, such as
those predicted in the chosen inflaton model, and manifest in the matter
power spectrum as modulations around intermediate field values. At early
times (high redshifts $z=1100,1300,1500$), corresponding to
post-recombination epochs, the power spectrum is suppressed due to the
limited growth of structures, as encoded in the growth factor $D(z)$. In
this regime, the shape of the spectrum primarily reflects primordial
conditions and transfer effects. As the universe evolves toward lower
redshifts $(z=70,50,30)$, the gravitational growth of matter overdensities
enhances the amplitude of $P(k,z)$, particularly for modes that exited the
horizon earlier (larger $\phi _{k}$). The impact of increasing $b$ is seen
as more pronounced and persistent oscillatory structures in the power
spectrum, especially for intermediate $\phi _{k}$ values, indicating a
direct mapping between inflaton dynamics during inflation and the late-time
structure of the universe. The results emphasize how the inflaton field at
horizon crossing not only sets the scale of each Fourier mode but also
determines how specific features from inflation can imprint on observable
large-scale structures, modulated over cosmic time by redshift-dependent
growth.

\section{Gravitational Collapse of a dark matter halos}

\label{sec3}

\subsection{Gravitational collapse of Overdense regions}

Observations of the cosmic microwave background \cite{Planck 2018, Bennett
1996, Planck 2020} show that the universe at recombination was highly
homogeneous, with initial density and gravitational potential fluctuations
of order $\Delta \rho /\rho \sim 10^{-5}$ \cite{Barkana 2001}. These
primordial perturbations, likely generated during inflation \cite{Guth 1981,
Starobinsky 1982}, grew via gravitational instability \cite{Peebles 1989}
over cosmic time, eventually forming the galaxies and large-scale structure
observed today \cite{Springel 2005}.

In cosmological modeling, the fluid description provides a powerful
framework for understanding large-scale structure formation \cite{Peebles
1982, Mo 2010}. This approach treats both dark matter and baryonic matter as
continuous fluids governed by gravity dynamics \cite{Bertschinger 1995}. For
cold dark matter, this approximation remains valid because its extremely
weak interactions mean the particles' dynamics are dominated by
gravitational forces on cosmological scales \cite{Dodelson 2003, Bullock
2017}. Similarly, baryons can be described as a collisional fluid until
nonlinear processes become important \cite{Cen 1992}. The fluid
approximation accurately models CDM dynamics until shell-crossing occurs 
\cite{Zel'dovich 1970, Hahn 2016}, when particle streams intersect. This
transition to nonlinearity requires switching to N-body methods \cite%
{Springel 2005}. Baryons behave as a pressureless fluid when thermal
pressure is negligible $\left( T\leq 10^{4}K\right) $ \cite{Barkana 2001},
but develop shocks during nonlinear collapse \cite{Ryu 1993}. For linear
perturbations ($\delta \ll 1$), the evolution equation simplifies to \cite%
{Kolb 1990}:%
\begin{equation}
\frac{\partial ^{2}\delta }{\partial t^{2}}+2H\frac{\partial \delta }{%
\partial t}=4\pi G\bar{\rho}\delta ,
\end{equation}%
here $\delta $\ is the density contrast, $H$ is the Hubble parameter, $G$ is
the gravitational constant, and $\bar{\rho}$ is the matter density,\ this
admits growing and decaying mode solutions \cite{Eisenstein 1998}, with the
growing mode dominating structure formation from Gaussian initial conditions 
\cite{Barkana 2001}.

To study structure formation across different mass scales, we analyze the
smoothed density perturbation field. By convolving the density contrast $%
\delta (x)$ with a normalized window function $W(y)$, we obtain a
Gaussian-distributed field with zero mean. For a spherical top-hat filter
(where $W=1$ within radius $R$), this directly measures mass fluctuations in
spheres of radius $R$ \cite{Barkana 2001}.%
\begin{equation*}
\sigma ^{2}(M)=\sigma ^{2}(R)=\int_{0}^{\infty }\frac{dk}{2\pi ^{2}}k^{2}%
\mathcal{P}(k,z)\left[ \frac{3j_{1}\left( kR\right) }{kR}\right] ^{2},
\end{equation*}%
where $j_{1}\left( x\right) =\left( \sin x-x\cos x\right) /x^{2}.$\ The mass
variance function $\sigma $ serves as a fundamental quantity in predicting
the statistical distribution of collapsed cosmological structures.

\begin{figure}[h]
\centering
\includegraphics[width=0.99\textwidth]{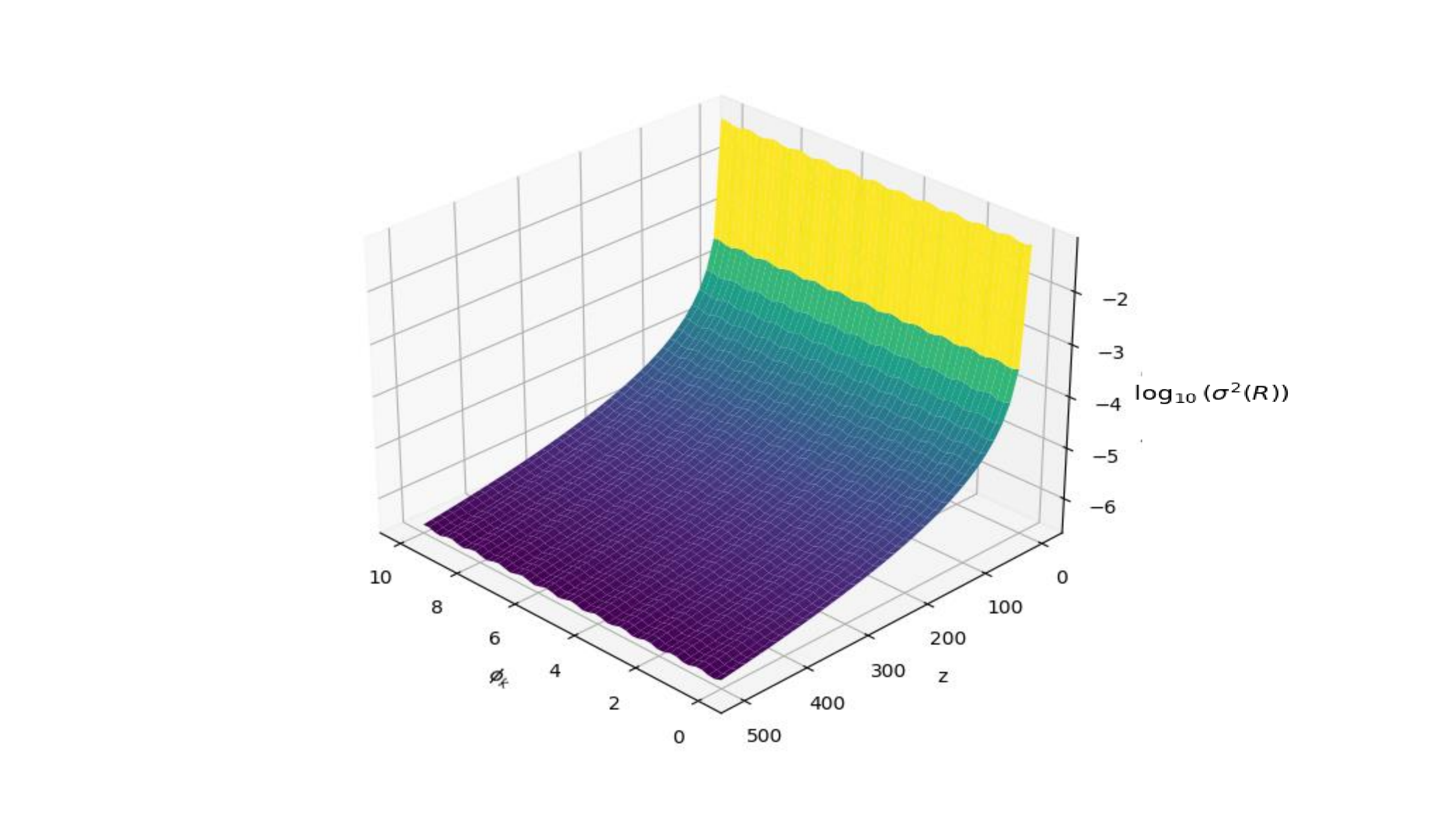} 
% Adjust width as needed
% Label for referencing the figure
\caption{ The variance of overdense regions collapse as functions of field
value at the horizon crossing and the redshift.}
\label{fig:3}
\end{figure}

Fig. \ref{fig:3} computes the variance of matter density fluctuations, $%
\sigma ^{2}(R)$ across cosmic time as a function of the inflaton field value
at horizon crossing $\phi _{k}$ , and the redshift $z$, for a fixed physical
radius that corresponds to typical scales of collapsing overdense regions at
the end of the dark ages, where the first structures such as minihalos and
primordial galaxies began to form, the choice of minihalo masses in the
range $10^{5}M_{\odot }$ to $10^{8}M_{\odot }$ is grounded in both
theoretical predictions and high-resolution cosmological simulations that
probe the onset of structure formation during the post-dark age era, in our
simularion we chose the radius of forming overdense regions corresponding to
the indicated masse ranges as $R\approx 0.5kpc$. The variance is obtained by
integrating the matter power spectrum $\mathcal{P}(k,z)$, weighted by a
top-hat window function in Fourier space, over a range of comoving
wavenumbers $k$. The power spectrum is derived from a primordial
inflationary spectrum modulated by an oscillatory term $\delta n_{s}\cos
\left( \phi _{k}/f\right) $. These oscillations in $\phi _{k}$ encode
spatially varying imprints left by inflation, leading to localized
enhancements and suppression in the amplitude of primordial fluctuations. As
a result, the spatial distribution of matter perturbations becomes
non-uniform, causing certain regions to collapse earlier or more efficiently
than others. This modulation plays a crucial role in shaping the small-scale
structure of the universe and governs the statistical abundance and spatial
distribution of overdensities that eventually collapse into halos. The
resulting plot of $\sigma ^{2}(R)$ \ shows that as the universe evolves
toward redshifts $z\sim 10-30$ during the final stages of the dark ages, the
logarithm of the variance squared, exceeds $-2$, indicating a significant
enhancement in matter fluctuations. This increase in variance greatly boosts
the likelihood of structure formation, making the collapse of overdense
regions far more probable compared to earlier epochs at higher redshifts,
where the variance remains suppressed and closely tied to the nearly uniform
conditions prevailing near the time of the $CMB$.

\subsection{Dark Matter Halos Abundance}

The tiny ripples in matter density we see in the CMB \cite{Planck 2020}
slowly grow over billions of years due to gravity's pull \cite{Peebles 2020}%
. Eventually, these small fluctuations become so dense that their collapse
can no longer be described by simple linear math and requires more complex
nonlinear physics. While exact solutions for how dark matter halos form are
only possible for perfectly symmetric cases \cite{Padmanabhan 1993}, most
real-world structure formation can be accurately modeled using Newtonian
gravity as long as we're looking at regions much smaller than the observable
universe \cite{Ellis 1999}. The simplest useful model imagines a perfectly
spherical blob of slightly higher density \cite{Gunn 1972}. Even though this
"spherical collapse" model ignores messy real-world details, it surprisingly
predicts many observed properties of dark matter halos, like their abundance
and distribution \cite{Wechsler 2018}.

Initially, small density fluctuations grow gently according to linear
theory's predictions ($\delta _{L}=\delta _{i}D\left( t\right) /D\left(
t_{i}\right) $), as established in \cite{Peebles 2020}. However, as
gravitational amplification becomes nonlinear \cite{Press 1974, Press 1974p}%
, bound regions (with negative total Newtonian energy) undergo a
characteristic evolution: initial expansion, turnaround at maximum radius,
and eventual collapse \cite{Gunn 1972}. The critical linear overdensity at
collapse, $\delta _{L}=1.686$, was first derived for Einstein-de Sitter
cosmology \cite{Peebles 2020} and shows only weak dependence on cosmological
parameters ($\Omega _{m},$ $\Omega _{\Lambda }$) as demonstrated by \cite%
{Eke 1996}. This threshold provides a powerful predictive tool - any region
whose linearly extrapolated present density contrast reaches $\delta
_{c}(z)=1.686/D(z)$, where $D(z)$ is the growth function \cite{Carroll 1992}%
, will collapse by redshift $z$. This framework, validated by N-body
simulations \cite{Jenkins 2001}, remains fundamental for predicting
structure formation timelines \cite{Wechsler 2018}.

A crucial test of structure formation theories is predicting the halo mass
function is the number density of dark matter halos as a function of mass
and redshift. This prediction links dark matter to observable galaxies and
clusters, enabling cosmological constraints. While simulations like
Millennium \cite{Springel 2005} measure halo abundances numerically,
analytical models \cite{Press 1974, Press 1974p} are built upon the concepts
of a Gaussian random field for density perturbations, linear gravitational
evolution, and spherical collapse. To compute the halo abundance at redshift 
$z$, we use $\delta _{M}${}, the density field smoothed on mass scale M.
Since $\delta _{M}$ is Gaussian with zero mean and standard deviation $%
\sigma (M)$, the probability that $\delta _{M}>\delta $ is \cite{Barkana
2001},%
\begin{equation}
\int_{\delta }^{\infty }d\delta _{M}\frac{1}{\sqrt{2\pi }\sigma (M)}e^{ -%
\frac{\delta _{M}^{2}}{2\sigma ^{2}(M)}} =\frac{1}{2} erfc\left( \frac{%
\delta }{\sqrt{2}\sigma }\right) .
\end{equation}

This probability is identified with the fraction of dark matter particles in
halos of mass greater than or equal to $M$ that have collapsed by a redshift 
$z$. Adjustments can be made by setting the density threshold $\delta $ is
set to $\delta _{c}(z)$ then the result is multiplied by 2 to account for
all particles \cite{Barkana 2001}. This yields the final expression for the
mass fraction in halos above $M$ at redshift $z$.%
\begin{equation}
\mathcal{F}(>\left. M\right\vert z)=erfc\left( \frac{\delta _{c}(z)}{\sqrt{2}%
\sigma }\right) ,
\end{equation}%
with the notation '$>\left. M\right\vert z$' means the fraction of dark
matter particles that are part of halos of mass $m>M$ at redshift $z$. The
factor of $2$ corrects for counting only positive fluctuations of $\delta
_{M}$. In Ref. \cite{Bond 1991} rederived this factor through the
"cloud-in-cloud" problem, showing that regions with $\delta _{M}<${}$\delta
_{c}(z)$ can still reside within larger collapsed regions $(M_{L}>M)$, thus
contributing to halos of greater mass. This naturally leads to a factor of $%
2 $ correction, increasing the fraction of particles in halos above mass $M$.

\begin{figure}[h]
\centering
\includegraphics[width=0.7\textwidth]{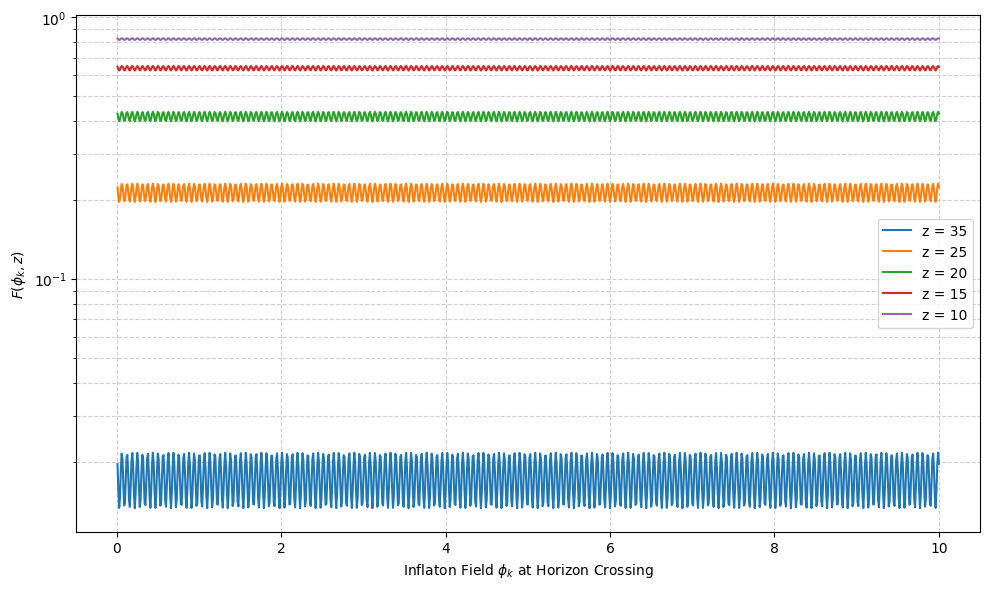} % Adjust width as needed
% Label for referencing the figure
\caption{ The mass fraction as functions of $\protect\phi _{k}$ for
different values of the redshift $z$. }
\label{fig:4}
\end{figure}

Fig. \ref{fig:4} visualizes the collapse fraction $\mathcal{F}(>\left.
M\right\vert z)$, which quantifies the probability that matter collapses
into bound structures at a given redshift $z$ and as a function of the
inflaton field value $\phi _{k}$ at horizon crossing. The collapse fraction
is computed using the complementary error function, applied to the ratio of
the critical density threshold for collapse, to the square root of twice the
variance of the matter density field. The variance is obtained by
integrating the matter power spectrum $\mathcal{P}(k,z)$, weighted by a
spherical top-hat window function over a radius around $R\approx 0.5kpc$,
which represents the physical size of early collapsing halos during the end
of the dark ages. The matter power spectrum itself includes both transfer
and growth effects and originates from a primordial curvature power spectrum
modulated by inflationary oscillations. These oscillations reflect periodic
features in the inflaton potential, such as those predicted by
axion-monodromy inflation, and lead to spatial modulations in early density
perturbations. By plotting $\mathcal{F}(>\left. M\right\vert z)$ across a
range of redshifts from $10$ to $35$, the plot reveals how the likelihood of
gravitational collapse evolves over time and how inflationary field dynamics
influence the spatial pattern of structure formation. The use of a
logarithmic scale highlights the variation in collapse probability across
different values of $\phi _{k}$ , emphasizing the role of inflation-induced
oscillations in seeding the earliest nonlinear structures in the universe.

\section{First Star Formation}

\label{sec4}

\subsection{Primordial Gas Collapse}

The $\Lambda $CDM model, precisely calibrated by WMAP \cite{Larson 2011} and
Planck \cite{Planck 2018, Planck 2020}, establishes our modern understanding
of early star formation. In this hierarchical structure formation paradigm,
the first stars likely emerged in dark matter minihalos ($\sim
10^{5}-10^{6}M_{\odot }$) at $z\sim 20-30$. Our understanding of the
earliest cosmic structures centers on dark matter minihalos, which formed
from amplified quantum fluctuations in the primordial density field. These
overdensities grew via gravitational instability \cite{Peebles 1982} until
reaching turnaround and collapsing out of the Hubble flow, ultimately
achieving virial equilibrium while accreting pristine baryonic gas.

The spherical collapse model \cite{Gunn 1972, Tegmark 1997} provides an
analytic framework for this process given as,%
\begin{equation}
R_{vir}\simeq 200pc\left( \frac{M_{h}}{10^{6}M_{\odot }}\right) ^{1/3}\left( 
\frac{1+z}{10}\right) ^{-1}\left( \frac{\Delta _{c}}{200}\right) ^{-1/3},
\end{equation}%
here $M_{h}$ is the total halo mass, $\Delta _{c}=\rho _{vir}/\rho _{b}${}
represents the overdensity once virialization is (almost) fully achieved,
with $\rho _{vir}\simeq 200\rho _{b}$\ and $\rho _{b}\simeq 2.5\times
10^{-30}\left( 1+z\right) ^{3}g~cm^{-3}$ are the virial and the background
densities. The gas heats up during collapse through adiabatic compression or
shock heating, reaching a virial temperature linked to the virial velocity
of dark matter particles, given by \cite{Barkana 2001},%
\begin{equation}
T_{vir}\simeq 2\times 10^{3}K\left( \frac{M_{h}}{10^{6}M_{\odot }}\right)
^{2/3}\left( \frac{1+z}{20}\right) .
\end{equation}%
Gas in minihalos generally stays below the $\sim 10^{4}K$ cooling threshold,
preventing further collapse and star formation, and instead remains in
hydrostatic equilibrium following the dark matter profile.

The thermodynamics of primordial gas is mainly regulated by $H_{2}$ cooling,
which depends on non-equilibrium gas-phase chemistry due to the absence of
dust grains \cite{Galli 1998}. Because $H_{2}$ lacks a permanent dipole
moment, it cannot easily radiate excess energy during formation, making
direct $H-H$ collisions inefficient \cite{Gould 1963, Gould I 1963}. In the
interstellar medium, dust grains catalyze $H_{2}$ formation by absorbing
this energy \cite{Hollenbach 1979}, but such a process was unavailable in
the early universe \cite{Tielens 1985}. Instead, the dominant formation
pathway involves free electrons as catalysts: $H+e^{-}\rightarrow
H^{-}+\gamma $, followed by $H+e^{-}\rightarrow H^{-}+\gamma $. An
alternative channel involving becomes relevant at very high redshifts $z>100$
when CMB photons can destroy $H^{-}$ but not $H_{2}^{+}$. As electrons
recombine and vanish, $H_{2}$ formation declines. Simulations show that the
resulting $H_{2}$ abundance increases with the virial temperature of halos,
following $f_{H_{2}}\propto T_{vir}^{1.5}$, implying that primordial gas
must first heat up through collapse before it can cool efficiently via
molecular hydrogen. 
\begin{figure}[h]
\centering
\includegraphics[width=0.859\textwidth]{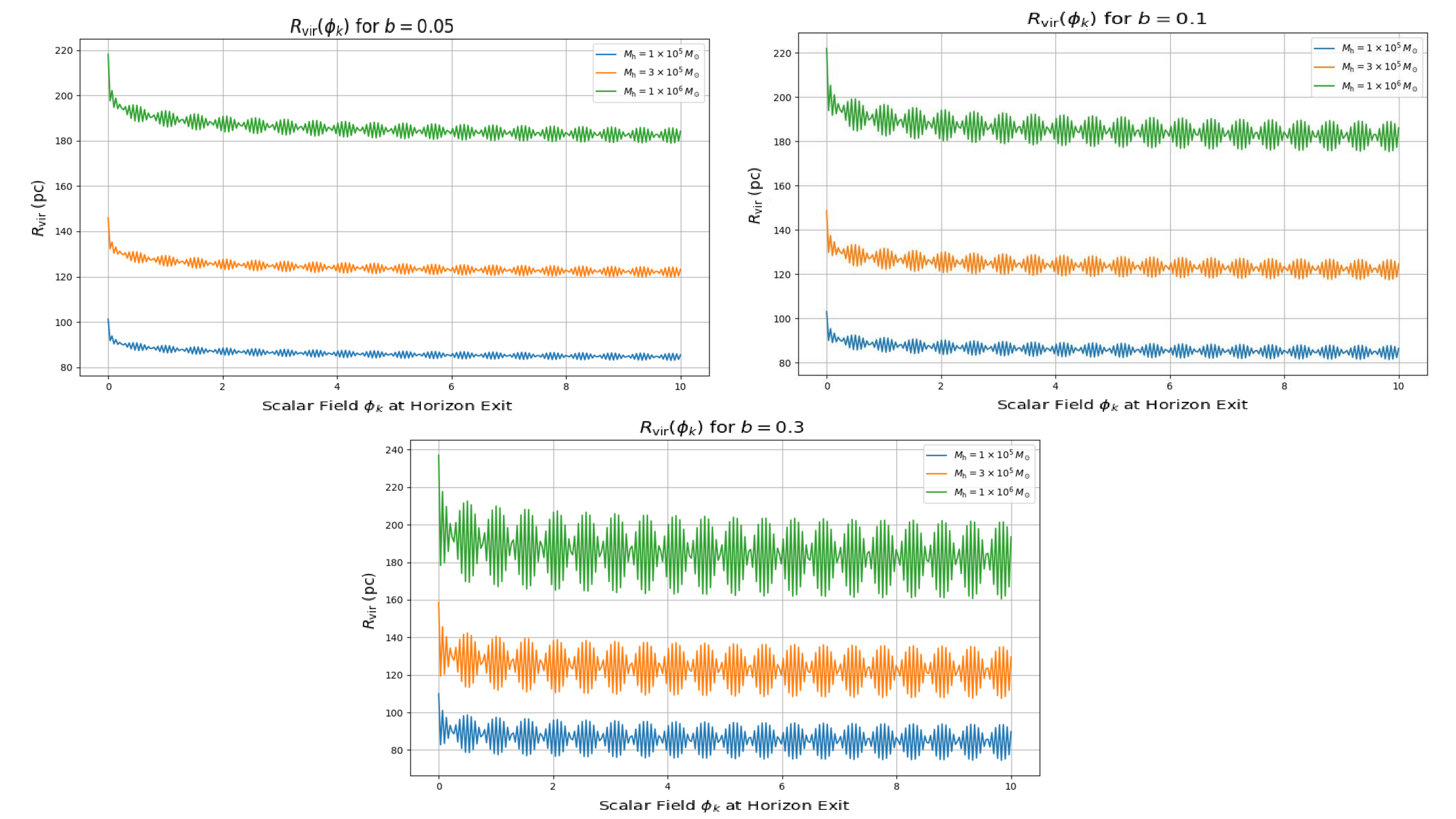} 
% Adjust width as needed
% Label for referencing the figure
\caption{ Virial Radius $R_{\mathrm{vir}}$ vs. $\protect\phi _{k}$ for
Various $M_{h}$.}
\label{fig:5}
\end{figure}

Fig. \ref{fig:5} investigates how the virial radius $R_{\mathrm{vir}}$ of
dark matter halos responds to changes in the inflaton field value at horizon
crossing $\phi _{k}$ for different values of $M_{h}$. The parameter $b$
modulates the strength of these oscillations, affecting the amount of
small-scale power available to seed structure formation. For each value of $%
b $, the code computes the growth factor $D(k,z)\propto \mathcal{P}%
^{1/2}(k,z)/\Delta _{\mathcal{R}}(k)$, which quantifies how density
perturbations grow after horizon crossing, and uses it to calculate the
virial radius of halos of different masses. The virial radius, representing
the equilibrium size of a collapsed halo, depends on both the halo mass and
the growth of the corresponding fluctuation mode. Each subplot corresponds
to a different value of $b$: $0.05$, $0.1$, and $0.3$, showing that as $b$
increases, the amplitude of oscillations in the primordial spectrum grows,
leading to increasingly pronounced variations in the virial radius across
scales. This behavior highlights the sensitivity of cosmic structure to the
detailed physics of the inflationary epoch, showing how early-universe
dynamics can imprint observable signatures on halo formation. 
\begin{figure}[h]
\centering
\includegraphics[width=0.99\textwidth]{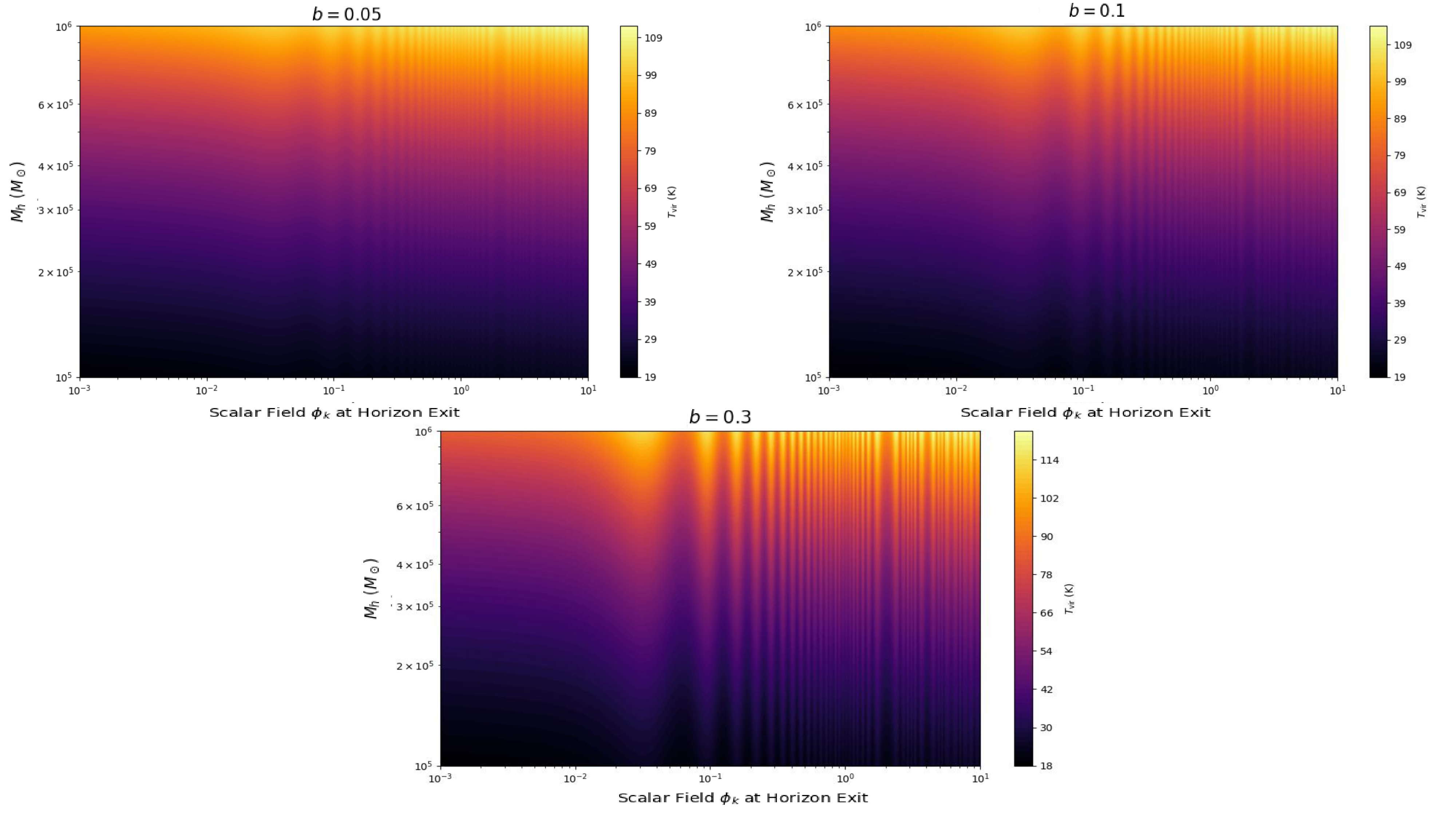} 
% Adjust width as needed
% Label for referencing the figure
\caption{ Virial Temperature $T_{vir}$ vs. $\protect\phi _{k}$ for different
values of $b$.}
\label{fig:6}
\end{figure}
Fig. \ref{fig:6} the virial temperature $T_{vir}$ of dark matter halos as
functions of $\phi _{k}$ and the halo mass $M_{h}$, for different values of
the oscillatory amplitude parameter $b$, which arises from the
axion-monodromy inflationary model. The virial temperature is a critical
quantity in astrophysics, representing the thermal state of gas after it
collapses and virializes within a dark matter halo, often determining
whether the gas can cool and form stars. The figure incorporates the effects
of primordial power spectrum oscillations on structure formation. This
structure reflects deviations from perfect scale-invariance due to
underlying inflaton dynamics during the early universe. The growth factor
derived from this spectrum, quantifies the amplification of density
perturbations post-inflation. Since $T_{vir}\propto M_{h}^{2/3}${}$/D(z)$,
both the halo mass and the inflationary physics directly impact the
temperature achieved during collapse. We computes the virial temperatures
across a logarithmic range of $\phi _{k}$ and $M_{h}$, and generates a plot
for each value of $b$. The results reflect the effect of the proportionality
of the halo mass to the virial temperature the increasing halo mass to
values above $M_{h}\propto 10^{5}M_{\odot }$, while increasing the $b$
values as $b\geq 0.05$ shows the effect of oscillations when increasing the
the inflaton field parameter at the horizon crossing $\phi _{k}$.

\subsection{Formation process of early stars}

During the collapse of primordial gas in a minihalo, three density-dependent
phases emerge, each with distinct thermal and chemical processes \cite{Abel
2002, Bromm 2004}. Initially thought to be simpler than present-day star
formation, simulations now reveal increasing complexity \cite{Yoshida 2006}.
This shift marks a maturation of the field beyond its early exploratory
phase \cite{Greif 2011, Clark 2011}. At densities below $10^{8}cm^{-3}$,
primordial gas remains mostly atomic with a small fraction of molecular
hydrogen $n\left[ H_{2}\right] /n\sim 10^{-3}$, formed via gas-phase
reactions catalyzed by residual free electrons from recombination or
ionization events \cite{Abel 2002, Galli 1998}. This small amount of $H_{2}$
enables cooling down to $\sim 200K$ at $n\sim 10^{4}cm^{-3}$, where $H_{2}$
transitions from non-$LTE$ to $LTE$ populations known as the "loitering
state". At this stage, cooling becomes less efficient, and the
characteristic density and temperature imprint the mass scale of Population
III pre-stellar cores, described by the Bonnor-Ebert mass \cite{Bromm 2002},%
\begin{equation}
M_{J}\simeq 500M_{\odot }\left( \frac{T}{200K}\right) ^{3/2}\left( \frac{n}{%
10^{4}cm^{-3}}\right) ^{-1/2}.
\end{equation}%
This pre-stellar core is at the verge of gravitational runaway collapse, and
it is the immediate progenitor of a Pop III star \cite{Barkana 2001}.

At densities above $10^{8}cm^{-3}$, primordial gas becomes fully molecular
through three-body reactions, greatly enhancing cooling by a factor of $\sim
10^{3}$ \cite{Palla 1983}. However, this cooling is moderated by heating
from $H_{2}$ formation, leading to a near-isothermal collapse at $\sim 1000K$%
. The equation of state also softens due to molecular degrees of freedom.
Significant uncertainties remain in the reaction rates \cite{Wise 2011}, and
experimental constraints are limited due to the neutral nature of the
involved particles \cite{Kreckel 2010}. At densities exceeding $%
n>10^{12}cm^{-3}$, the ro-vibrational lines of molecular hydrogen become
increasingly optically thick, making radiative transfer more complex. To
address this, recent simulations have advanced by employing an escape
probability approach in conjunction with the Sobolev approximation \cite%
{Clark 2011, Greif 2011, Greif 2012}. The method utilizes the following
expression for the escape probability:

\begin{equation}
\beta _{esc}=\frac{1-\exp \left( -\tau \right) }{\tau },
\end{equation}

The Sobolev approximation estimates the line optical depth as $\tau
=k_{lu}L_{char}$, with $L_{char}\approx v_{th}/\left\vert
dV_{r}/dr\right\vert .$\ This allows photons to escape from regions with
velocity gradients once Doppler shifts move them out of resonance.
Simulations using this method closely match \textit{1D}
radiation-hydrodynamic results, validating the approach \cite{Omukai 1998}.

By the end of the initial collapse, a small protostellar core forms at the
center of the minihalo. Recent high-resolution cosmological simulations have
enabled detailed modeling of this early stage \cite{Greif 2012, Stacy 2013}.
A key question in interpreting these results is how to define the
protostar's size. One approach is to use the photospheric radius, where the
optical depth $\tau \sim 1$, or equivalently, where the escape probability
reaches $\beta _{esc}=1-e^{-1}\approx 0.63$. Another method is to define the
hydrostatic radius as the region where the radial velocity approaches zero,
indicating equilibrium. In practice, the hydrostatic radius is smaller than
the photospheric radius, which is initially around $R_{p}\sim
100-200R_{\odot }$ \cite{Stahler 1986}. 
\begin{figure}[h]
\centering
\includegraphics[width=0.99\textwidth]{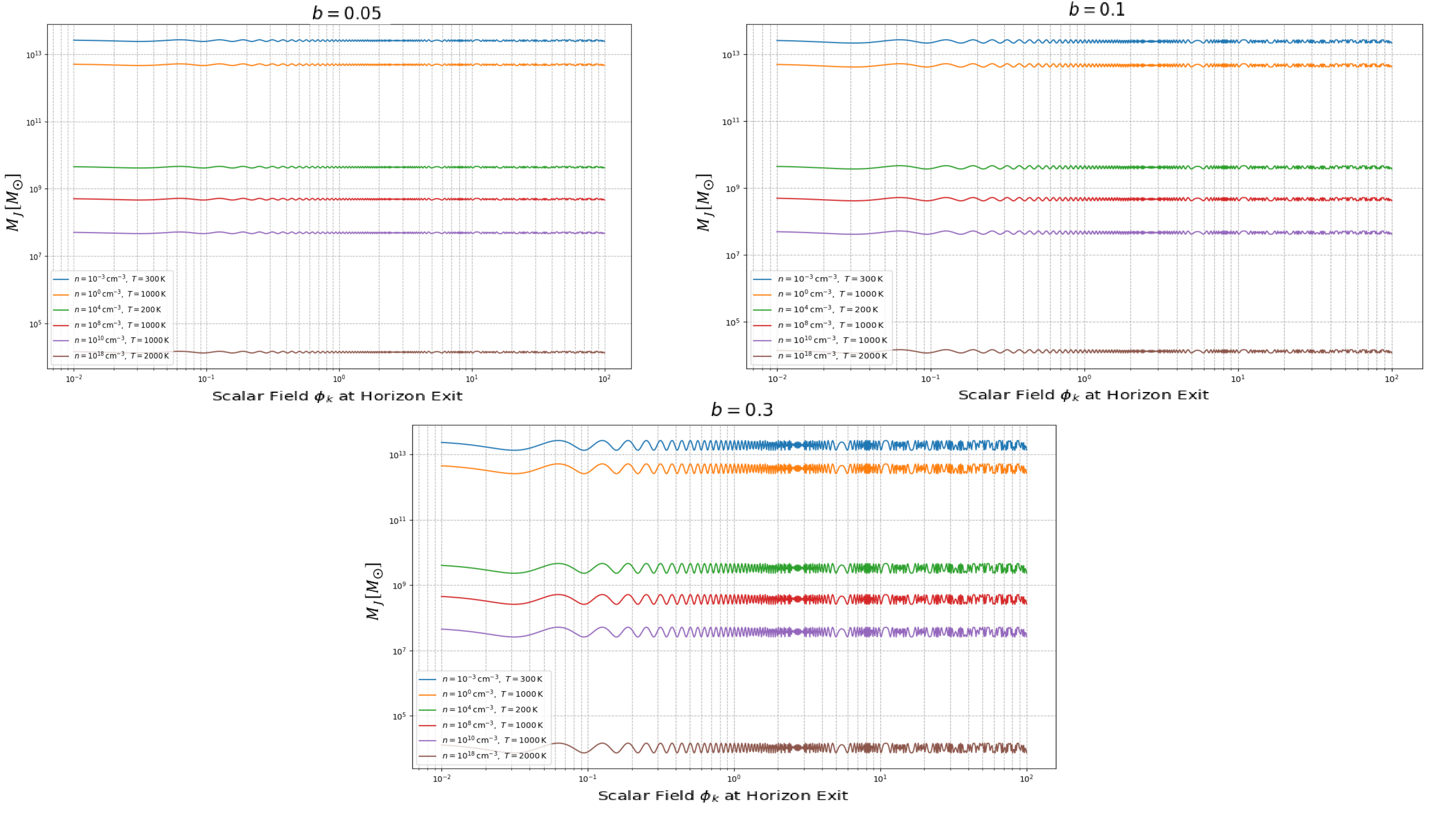} 
% Adjust width as needed
% Label for referencing the figure
\caption{ The Bonnor-Ebert Mass vs. The Inflation Field $\protect\phi _{k}$.}
\label{fig:7}
\end{figure}

Fig. \ref{fig:7} illustrates how the Bonnor-Ebert mass $M_{J}$, which
represents the minimum mass required for a gas cloud to undergo
gravitational collapse, varies with the inflationary field scale $\phi _{k}$
for different thermal phases of primordial gas. The analysis incorporates
inflationary oscillations in the primordial power spectrum, modeled through
a modulation term whose amplitude is controlled by the parameter $b$. For
each case $b=0.05$, $b=0.1$, and $b=0.3$, the power spectrum is modified
accordingly, influencing the growth factor $D(k,z)$ and, in turn, the mass
thresholds for collapse. When $b=0.05$, the oscillations are weak, and the
Bonnor--Ebert mass varies smoothly with $\phi _{k}$, closely resembling
predictions from standard $\Lambda $CDM cosmology. At $b=0.1$, moderate
oscillations introduce noticeable, periodic modulations in $M_{J}$,
indicating that certain inflationary scales could preferentially enhance or
suppress structure formation. In the case of strong oscillations $b=0.3$,
the mass thresholds exhibit significant fluctuations, suggesting that the
conditions for collapse vary drastically across different scales,
potentially imprinting signatures in the initial mass function of stars or
the early distribution of dark matter halos. Overall, the plot connects
inflationary physics to astrophysical processes in the early universe,
showing the primordial fluctuations imprints on the scales at which
structure first forms.

\section{Inflationary Signatures in PBH Formation \& Galaxy Properties}

\label{sec5}

\subsection{ Primordial Black Holes as Seeds for Early Galaxies}

The formation of massive primordial black holes as seeds for Active Galactic
Nuclei (AGN) was first proposed in \cite{Khlopov 2002, Khlopov 2005, Khlopov
2007, Khlopov 2010} in connection with inflationary dynamics of
Pseudo-Nambu-Goldstone scalar field possessing two distinct degenerated
vacuum states. As inflation ends, these vacuum states are unevenly
populated. Dominant part of the Universe possess one of them, but some
regions possess the second one. During the subsequent cooling of the
universe, the expansion rate slows, allowing the scalar field to evolve
freely toward the minima of its potential. This leads to the emergence of
regions where the second type of vacuum is enclosed within regions of the
first one. These "islands" of the second vacuum are separated by domain
walls from the surrounding regions of the first type of vacuum. When such
closed walls enter horizon they can collapse and produce black holes, with
the resulting PBH mass tied to the mass of the original vacuum wall. In
general, PBHs can also form from the gravitational collapse of
large-amplitude density fluctuations in the early universe \cite{Barbieri
2024}. Observational constraints on PBH abundance, derived from various
astrophysical probes \cite{Sanchez 1997, Green 1997}, suggest they
contribute no more than $10-20\%$ of the total cosmic energy density. PBHs
are increasingly studied for their potential role in early galaxy formation,
dark matter, and the origin of supermassive black holes. Though they make up
only a small fraction of dark matter, PBHs can grow via baryonic accretion 
\cite{De Luca 2020, De Luca 2023} and are typically embedded in evolving
dark matter halos \cite{Yuan 2024}. Large PBHs may also seed density
fluctuations through the Poisson or seed effects \cite{Hoyle 1966, Carr
1983, Carr 1984}, with the mass scale $M_{B}$ and redshift $z_{B}$
determining the growth of these fluctuations during the matter-dominated era 
\cite{Carr 2018}.%
\begin{eqnarray}
M_{B} &=&\left\{ 
\begin{array}{c}
\frac{M_{PBH}f_{PBH}}{\left[ \left( 1+z_{B}\right) a_{eq}\right] ^{2}}%
~~~\left( Poisson~effect\right) , \\ 
\frac{M_{PBH}f_{PBH}}{\left( 1+z_{B}\right) a_{eq}}~~~\left(
Seed~effect\right) ,%
\end{array}%
\right.  \label{Seed} \\
M_{PBH} &=&M_{B}\frac{\left( 1+z\right) a_{eq}}{f_{PBH}}.
\end{eqnarray}

Here, $a_{eq}$ denotes the scale factor at matter--radiation equality. At
this stage, we assume $M_{B}\sim M_{halo}$ and the galaxy redshift $z\sim
z_{B}$. Additionally, we introduce the star formation efficiency $%
\varepsilon $, defined as:%
\begin{equation}
\varepsilon =\frac{M_{\ast }}{f_{b}M_{halo}},
\end{equation}%
where $M_{halo}$ represents the halo mass, and $f_{b}$ denotes the baryonic
fraction of the total matter content.

PBHs with masses above $10^{11}M_{\odot }${} are strongly constrained by
non-detections, except for rare cases like Phoenix A \cite{Brockamp 2016}.
Additional limits from $\mu $-distortion, $X$-ray binaries \cite{Inoue 2017}%
, Galactic center infall \cite{Carr 1999}, and large-scale structure \cite%
{Carr 2018} restrict PBHs with $M\gtrsim 10^{5}M_{\odot }${} to a dark
matter fraction $f_{PBH}\sim 10^{-4}-10^{-3}$ \cite{Su 2023}. Lyman-$\alpha $
forest data further exclude cases with $M_{PBH}f_{PBH}\gtrsim 170M_{\odot }$
and $f_{PBH}>0.05$ \cite{Murgia 2019}, challenging the viability of
supermassive PBHs under the Poisson effect. 
\begin{figure}[h]
\centering
\includegraphics[width=0.8\textwidth]{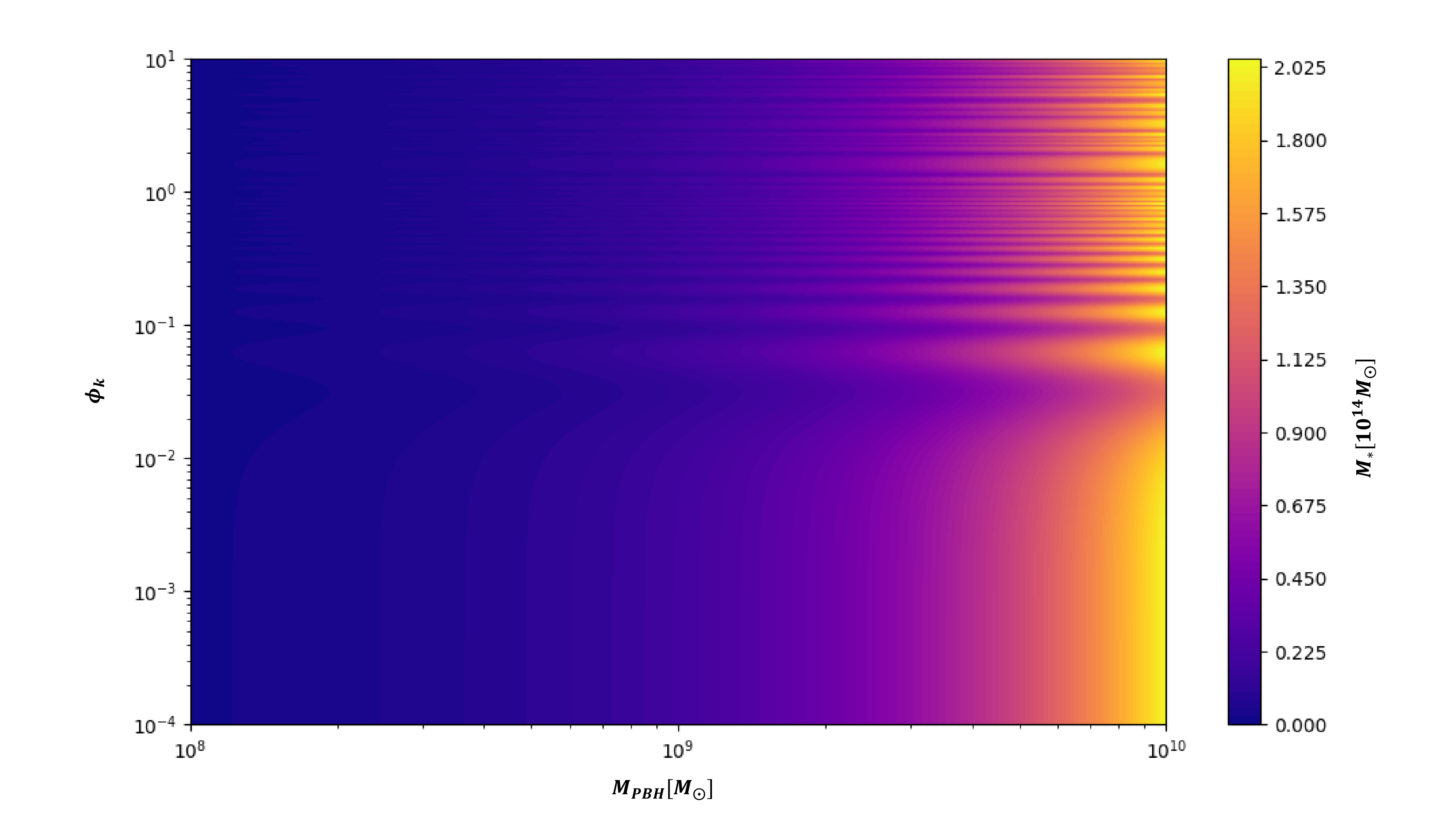} % Adjust width as needed
% Label for referencing the figure
\caption{Stellar Mass $M_\ast$ vs. $M_{\mathrm{PBH}}$ and $\protect\phi_k$.}
\label{fig:8}
\end{figure}
Fig. \ref{fig:8} illustrates how the stellar mass $M_{\ast }$ of primordial
galaxies depends on both the mass of PBHs, and the inflationary modulation
parameter $\phi _{k}$. In this scenario, PBHs act as early gravitational
seeds that drive the collapse of matter into halos, enabling the formation
of the first stars and galaxies. The stellar mass is calculated from the
halo mass, which is approximated by the baryonic mass $M_{B}$, derived from
the PBH seed effect given in Eq. (\ref{Seed}) The redshift $z$ at which
collapse occurs is not fixed but emerges from the inflationary power
spectrum $\Delta _{\mathcal{R}}^{2}(k)$, modulated by an oscillatory term.
This term encodes inflationary dynamics and induces scale-dependent
modulations in the collapse timing, thus imprinting early-universe physics
on galaxy-scale observables. The chosen PBH mass range $\left[ 10^{4},10^{10}%
\right] M_{\odot }$ is motivated by observationally inferred properties of
high-redshift galaxies. In particular, the galaxies \textit{JWST 35300 (G1)}
and \textit{JWST 38094 (G2)}, observed at redshifts $z\gtrsim 10$, exhibit
stellar masses of order $10^{10}M_{\odot }$, suggesting they may have formed
in massive halos potentially seeded by PBHs in this mass regime. These
galaxies, studied in works such as \cite{Boylan 2025, Labbe 2023}, challenge
standard $\Lambda $CDM formation timelines and lend credibility to
alternative early structure formation scenarios, including those involving
PBH seeding.

Supermassive black holes with masses $10^{6}-10^{7}M_{\odot }$ at redshifts $%
z>9$ challenge existing formation models due to uncertainties about seed
origins and rapid early growth \cite{Inayoshi 2020, Volonteri 2012}. GN-z11,
a galaxy at $z=10.6$, shows evidence of hosting an active SMBH, based on
JWST observations and the detection of high-ionization lines indicative of
an AGN \cite{Curti 2023, Bouwens 2010, Maiolino 2024, Terao 2022}. In what
follows, we apply accretion laws to primordial black holes, treating them as
potential seeds of the observed active galactic nuclei. Following the
approach of \cite{Valentini 2020}, we consider the bolometric luminosity,
represented by $L_{bol}=\varepsilon \dot{M}_{PBH}c^{2}.$ The Eddington ratio 
$\Gamma =L_{bol}/L_{Edd}${} quantifies the ratio between a quasar's
bolometric luminosity and its Eddington luminosity, assuming hydrostatic
equilibrium in a fully ionized hydrogen medium \cite{Eddington 2013}. It is
given by:%
\begin{equation}
L_{Edd}=\frac{4\pi M_{PBH}m_{p}c}{\sigma _{T}},
\end{equation}

In this study, we explore the Bondi--Hoyle--Lyttleton accretion model as a
framework for understanding PBH growth. In this scenario, PBHs accrete gas
from the surrounding medium at a rate determined by their mass, velocity,
and local gas density, with the process constrained by the Eddington limit,
which balances radiation pressure and gravity. Under favorable conditions
such as high gas densities PBHs may accrete at near- or even super-Eddington
rates, enabling rapid mass growth. By linking the Eddington accretion rate
and bolometric luminosity to the PBH mass and accretion efficiency $%
\varepsilon $, this model offers insight into how early PBHs could evolve
into the massive black holes observed at high redshifts. 
\begin{figure}[h]
\centering
\includegraphics[width=0.8\textwidth]{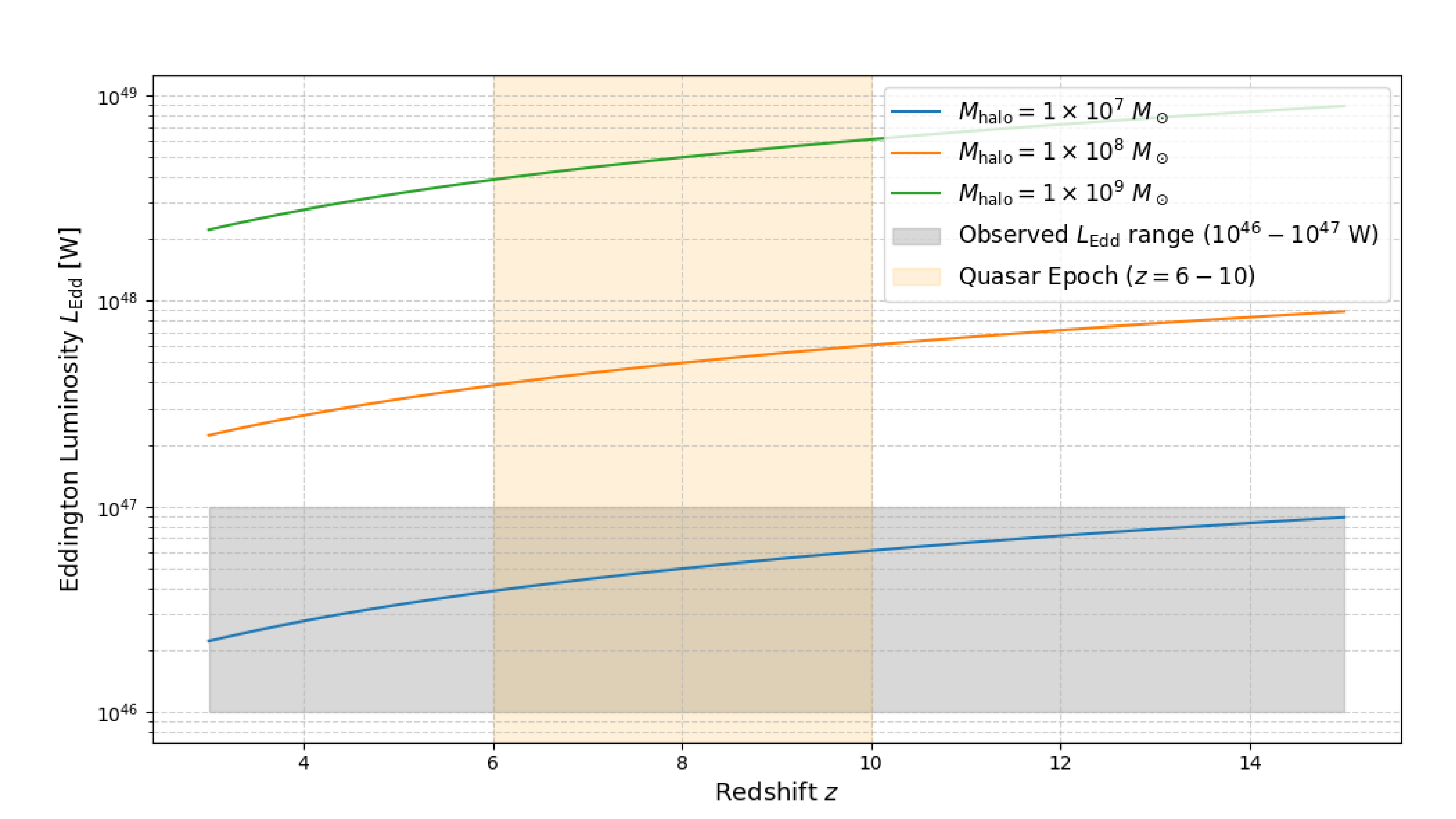} % Adjust width as needed
% Label for referencing the figure
\caption{Eddington Ratio as functions of the redshift for different Halo
masses.}
\label{fig:9}
\end{figure}
The high Eddington luminosities $L_{Edd}$ observed at early cosmic
times-especially for massive halos forming at redshifts $z\gtrsim 10$
provide a physical foundation for explaining the rapid growth of
supermassive black holes in the early universe. The Eddington luminosity
sets the maximum steady radiative output of an accreting black hole, above
which radiation pressure would counteract gravity and suppress further
accretion. It is directly proportional to the black hole mass MPBH, which,
in the seed scenario considered here, scales with halo mass and increases
significantly at high $z$ due to the stronger gravitational binding required
for collapse. In this framework, the seed mass of PBHs is given by Eq. (\ref%
{Seed}), indicating that PBHs embedded in halos forming at high redshift
must be more massive to trigger early star and galaxy formation. The
resulting Eddington luminosity $L_{Edd}$, can reach values above $10^{46}W$
for PBHs of masses around $10^{6}-10^{8}$ $M_{\odot }$, consistent with
observed luminosities of quasars at $z\sim 6-10$. Such high luminosities
imply that these black holes can accrete matter at nearly the Eddington rate
over extended periods, enabling the rapid growth required to form the
supermassive black holes with masses up to $\sim 10^{8}M_{\odot }$, as
inferred from JWST and ALMA observations \cite{Fan 2023, Pensabene 2021,
Walter 2022}. This rapid accretion scenario helps resolve the "quasar seed
problem"---how such massive black holes could form so early---by showing
that high-redshift PBHs embedded in overdense halos naturally yield large
Eddington luminosities, enabling efficient gas inflow and radiation
production. This supports recent interpretations that early SMBHs likely
underwent prolonged phases of near- or super-Eddington accretion, possibly
regulated by feedback processes, in order to reach their observed masses by $%
z\sim 7$. Fig. \ref{fig:9} shows the evolution of the Eddington luminosity
as a function of the redshift\ $z$, for different halo masses. Since the
Eddington luminosity is linearly proportional to the black hole mass, the
curves shows an increasing behavior of $L_{Edd}$. On the other hand, the
Eddington luminosity became lower when choosing higher values of the halo
mass. Moreover, the value $M_{halo}=10^{7}M_{\odot }$ reproduce the observed
value of the Eddington luminosity according to recent observations that also
aligns with its corresponding quasar epochs.

\subsection{Galactic Disk Formation and Structure}

The formation of disk galaxies was initially studied by \cite{Fall 1980}.
Later works by \cite{Dalcanton 1997, Mo 1998} examined disk size
distributions and compared them with observations. While various detailed
models exist, this study adopts a simplified approach: an exponential disk
embedded in a singular isothermal sphere halo. The analysis considers a halo
mass $M_{halo}$, virial radius $R_{d}$, total energy $E$ and angular
momentum $J$, characterized by the spin parameter $\lambda \equiv
J\left\vert E\right\vert ^{1/2}G^{-1}M_{halo}{}^{-5/2}.$ The spin parameter
provides a dimensionless measure of a halo's angular momentum. Assuming the
gas collapses into a rotationally supported disk within the dark matter
halo, and that the disk retains a fraction $m_{d}$ of the halo mass and $%
j_{d}$ of its angular momentum, the resulting disk scale radius can be
expressed as derived by \cite{Mo 1998}.

\begin{equation}
R_{d}=\frac{1}{\sqrt{2}}\left( \frac{j_{d}}{m_{d}}\right) \lambda R_{vir},
\end{equation}

Observations suggest that galactic disks retain a similar specific angular
momentum to their host halos, leading to the assumption $j_{d}/m_{d}=1$ \cite%
{Dalcanton 1997, Mo 1998}. However, the physical origin of this relation
remains uncertain. Numerous simulations \cite{Navarro 1991, Evrard 1994,
Navarro 1995, Tissera 1997} have shown that gas often loses angular momentum
to the dark matter halo during mergers, resulting in disks that are too
small compared to observations. One proposed solution is to delay gas
collapse via supernova feedback \cite{Eke 1996, Binney 2001, Efstathiou 2000}%
, although some simulations indicate that this mechanism may not
sufficiently prevent angular momentum loss \cite{Navarro 2000}. 
\begin{figure}[h]
\centering
\includegraphics[width=0.8\textwidth]{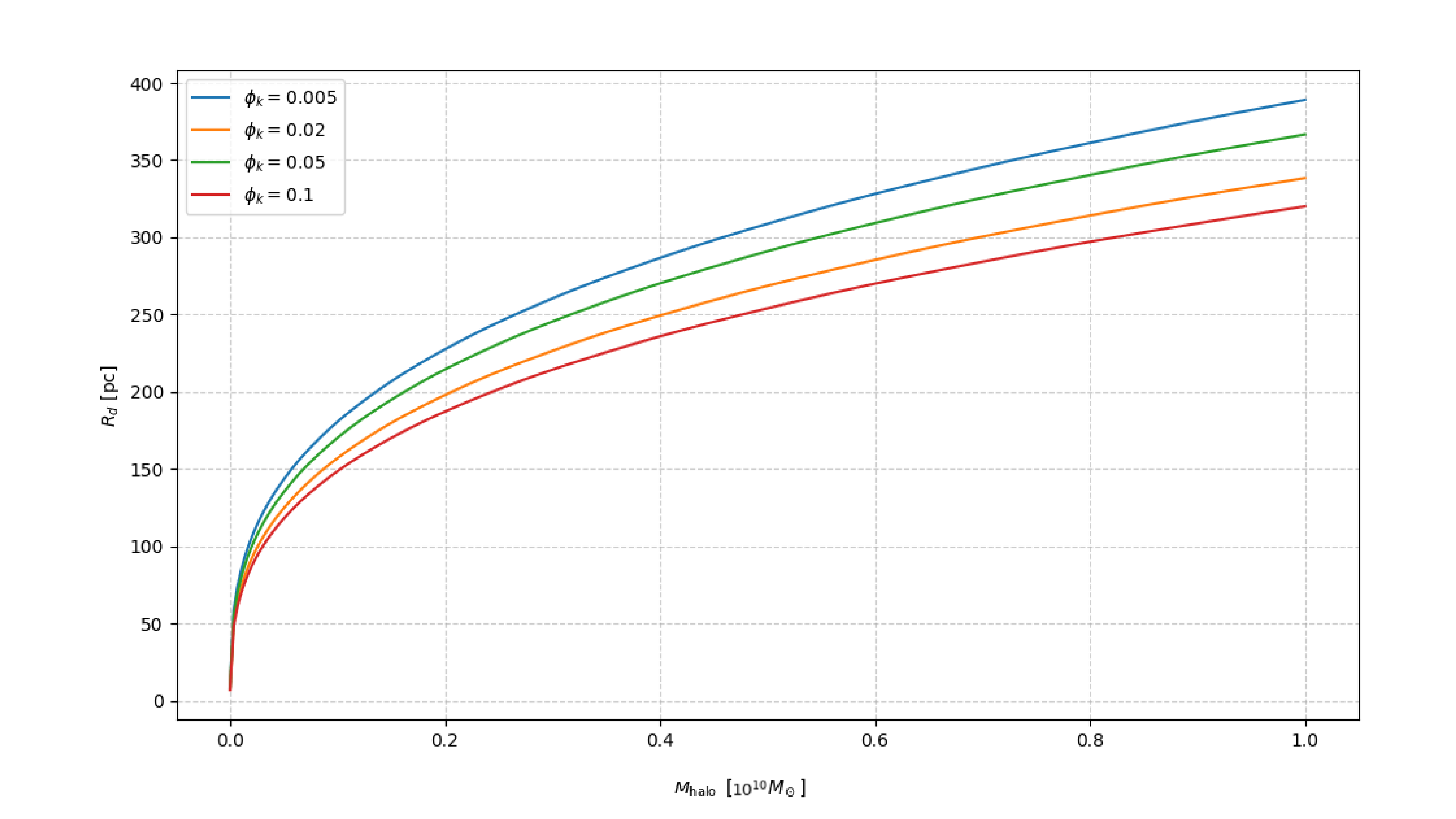} 
% Adjust width as needed
% Label for referencing the figure
\caption{Disk Radius $R_d$ as a Function of $M_{\mathrm{halo}}$ for Varying $%
\protect\phi_k$.}
\label{fig:10}
\end{figure}
The Fig. (\ref{fig:10}) models the disk radius $R_{d}$ of primordial
galaxies as a function of the host halo mass $M_{halo}$, incorporating the
influence of inflationary features encoded by the field parameter $\phi _{k}$%
. The theoretical basis follows the model developed by \cite{Mo 1998}, where
a galactic disk forms from baryonic gas that conserves angular momentum and
collapses within a dark matter halo modeled as a singular isothermal sphere.
The disk radius is determined by the virial radius $R_{vir}$, modulated by
the spin parameter $\lambda $ and the ratio $j_{d}/m_{d}\approx 1$, which
reflects the conservation of specific angular momentum between the halo and
the disk. The virial radius is dynamically computed as a function of
redshift, and the linear matter power spectrum that includes modifications
from inflationary physics via oscillations in the primordial power spectrum $%
\Delta _{\mathcal{R}}^{2}(k)$, parameterized by $b,~f,$ and $\phi _{k}$.
These introduce scale-dependent features through the spectral distortion
term $\delta n_{s}$, encoding the effects of primordial features on the
small-scale density perturbations relevant to early galaxy formation.

The resulting curve shows how the disk radius $R_{d}$ varies with $M_{halo}$
for different values of $\phi _{k}$, wherer the disk radius increases
moderately, but for more massive halos, $R_{d}$ grows more steeply. This
reflects the stronger sensitivity of disk size to halo mass at later stages
of collapse. Additionally, larger values of $\phi _{k}$ result in
systematically lower disk radii for a given halo mass, illustrating how
inflationary features specifically the phase of oscillations in $\Delta _{%
\mathcal{R}}^{2}(k)$ can impact the collapse history and angular momentum
retention in protogalactic structures.

\section{Conclusion}

\label{sec6}

This work has traced the evolution of cosmological structure from its
inflationary origins to the emergence of the first galaxies. Beginning with
the transfer function formalism, we showed how causal microphysics and
inflationary oscillations filter primordial perturbations, shaping the scale
dependence of the matter power spectrum. We then examined the collapse of
overdense regions, deriving the variance of density fluctuations and
demonstrating how the halo mass function emerges from linear theory
supplemented by the spherical collapse model. Extending this framework, we
analyzed the thermochemical processes governing the collapse of primordial
gas in minihalos, emphasizing the role of $H_{2}$ cooling and the sequence
of density-dependent phases that culminate in Population III star formation.
Finally, we considered the imprint of inflationary features on primordial
black holes and their potential to seed high-redshift galaxies, linking
accretion-driven black hole growth to observed stellar masses, disk radii,
and the properties of early quasars.

Taken together, these results underscore a consistent theme: signatures of
inflation are not confined to the cosmic microwave background, but propagate
through nonlinear gravitational collapse and baryonic physics into the
earliest astrophysical structures. Combination of inflationary scenario with
pattern of particle symmetry breaking can be reflected in creation of
topological defects and primordial black holes and nonlinear structures \cite%
{Khlopov 2010,PPNP}, as well as can explain \cite{Guo2024}not only JWST
evidence for early galaxy formation, but also for the origin of stochastic
gravitational wave background discovered by Pulsar Timing Arrays in nHz
range. Our present studies enrich possible observational signatures for such
phenomena. In particular, oscillatory features in the primordial spectrum
can modulate the abundance and timing of halo formation, alter the mass
scale of Population III stars, and influence the seeding and growth of
primordial black holes. With the advent of JWST and other high-resolution
probes, these theoretical connections provide a pathway for testing
inflationary models through galaxy and quasar observations at $z\gtrsim 10$.
Future refinements---incorporating radiative transfer, feedback, and
improved treatments of nonlinear collapse---will be essential for assessing
the robustness of these signatures. Nonetheless, the framework presented
here highlights the deep continuity between high-energy physics in the early
universe and the astrophysical phenomena that mark the dawn of structure
formation.

In this paper, we stated that oscillatory features in the primordial power
spectrum persist over cosmic time and can be probed by current and future
observations. These oscillatory features imprinted in the primordial power
spectrum propagate across cosmic time and may be detected by a range of
current and forthcoming observations. At high redshifts $(z\sim 30-70)$,
modulations of small-scale power affect the formation of minihalos and
Population III stars, leading to potentially detectable signatures in the
global 21-cm signal and its power spectrum. At later epochs $(z\gtrsim 10)$,
the modified halo mass function and virial properties predicted in our
framework may induce measurable signatures of these perturbations in the
abundance and clustering of early galaxies, accessible to current
observational technologies. On smaller scales, oscillatory enhancements of
the power spectrum can locally increase collapse probabilities, leaving
imprints in primordial black hole abundance constraints derived from CMB
spectral distortions and stochastic gravitational-wave backgrounds. A
complementary and more direct avenue is provided by cross-correlations among
early-Universe observables, which could enhance sensitivity to coherent
oscillatory patterns and help disentangle inflationary features from
astrophysical effects using existing and near-future observational data.

\section{Acknowledgements}

The work of M. K. was performed in Southern Federal University with
financial support of grant of Russian Science Foundation \^{a},,-- 25-07-IF.


\begin{thebibliography}{999}
\bibitem{Dodelson 2003} Dodelson, S. (2003, October). \textit{Coherent phase
argument for inflation}. In AIP Conference Proceedings (Vol. 689, No. 1, pp.
184-196). American Institute of Physics.

\bibitem{Mukhanov 2005} Mukhanov, V. (2005). \textit{Physical foundations of
cosmology}. Cambridge university press.

\bibitem{Bardeen 1987} Bardeen, J. M., Bond, J. R., \& Efstathiou, G.
(1987). \textit{Cosmic fluctuation spectra with large-scale power}.
Astrophysical Journal, Part 1 (ISSN 0004-637X), vol. 321, Oct. 1, 1987, p.
28-35. NSERC-supported research., 321, 28-35.

\bibitem{Eisenstein 1998} Eisenstein, D. J., \& Hu, W. (1998). \textit{%
Baryonic features in the matter transfer function}. The Astrophysical
Journal, 496(2), 605.

\bibitem{Peacock 1999} Peacock, J. A. (1999). \textit{Cosmological physics}.
Cambridge university press.

\bibitem{McAllister 2014} McAllister, L., Silverstein, E., Westphal, A., \&
Wrase, T. (2014).\textit{The powers of monodromy.} Journal of High Energy
Physics, 2014(9), 1-32.

\bibitem{Planck 2018} Aghanim, N., Akrami, Y., Ashdown, M., Aumont, J.,
Baccigalupi, C., Ballardini, M., ... \& Roudier, G. (2020). \textit{Planck
2018 results-VI}. Cosmological parameters. Astronomy \& Astrophysics, 641,
A6.

\bibitem{Guth 1981} Guth, A. H. (1981). \textit{Inflationary universe: A
possible solution to the horizon and flatness problems.} Physical Review D,
23(2), 347.

\bibitem{Starobinsky 1982} Starobinsky, A. A. (1982). \textit{Dynamics of
phase transition in the new inflationary universe scenario and generation of
perturbations}. Physics Letters B, 117(3-4), 175-178.

\bibitem{Peebles 1982} Peebles, P. J. E. (1982). \textit{Large-scale
background temperature and mass fluctuations due to scale-invariant primeval
perturbations}.

\bibitem{Mo 2010} Mo, H., Van den Bosch, F., \& White, S. (2010). \textit{%
Galaxy formation and evolution.} Cambridge University Press.

\bibitem{Bertschinger 1995} Bertschinger, E. (1995). \textit{Cosmological
dynamics}. arXiv preprint astro-ph/9503125.

\bibitem{Gunn 1972} Gunn, J. E., \& Gott III, J. R. (1972). \textit{On the
infall of matter into clusters of galaxies and some effects on their
evolution}. Astrophysical Journal, vol. 176, p. 1, 176, 1.

\bibitem{Padmanabhan 1993} Padmanabhan, T. (1993). \textit{Structure
formation in the universe}. Cambridge university press.

\bibitem{Press 1974} Press, W. H., \& Schechter, P. (1974). \textit{%
Formation of galaxies and clusters of galaxies by self-similar gravitational
condensation}. Astrophysical Journal, Vol. 187, pp. 425-438 (1974), 187,
425-438.

\bibitem{Springel 2005} Springel, V., White, S. D., Jenkins, A., Frenk, C.
S., Yoshida, N., Gao, L., ... \& Pearce, F. (2005). \textit{Simulations of
the formation, evolution and clustering of galaxies and quasars}. nature,
435(7042), 629-636.

\bibitem{Bond 1991} Bond, J. R., et al. "\textit{Excursion set mass
functions for hierarchical Gaussian fluctuations.}" Astrophysical Journal,
Part 1 (ISSN 0004-637X), vol. 379, Oct. 1, 1991, p. 440-460. Research
supported by NSERC, NASA, and University of California. 379 (1991): 440-460.

\bibitem{Peebles 1987} Peebles, P. J. E. (1987). \textit{Cosmic background
temperature anisotropy in a minimal isocurvature model for galaxy formation}%
. Astrophysical Journal, Part 2-Letters to the Editor (ISSN 0004-637X), vol.
315, April 15, 1987, p. L73-L76. NSF-supported research., 315, L73-L76.

\bibitem{Eke 1996} Eke, V. R., Cole, S., \& Frenk, C. S. (1996). \textit{%
Cluster evolution as a diagnostic for }$\Omega $\textit{.} Monthly Notices
of the Royal Astronomical Society, 282(1), 263-280.

\bibitem{Barkana 2001} Barkana, R., \& Loeb, A. (2001). \textit{In the
beginning: the first sources of light and the reionization of the universe}.
Physics reports, 349(2), 125-238.

\bibitem{Wechsler 2018} Wechsler, R. H., \& Tinker, J. L. (2018). \textit{%
The connection between galaxies and their dark matter halos}. Annual Review
of Astronomy and Astrophysics, 56(1), 435-487.

\bibitem{El Bourakadi 2023} El Bourakadi, K., Koussour, M., Otalora, G.,
Bennai, M., \& Ouali, T. (2023). \textit{Constant-roll and primordial black
holes in }$\mathit{f\ (Q,\ T)}$\textit{\ gravity}. Physics of the Dark
Universe, 41, 101246.

\bibitem{El Bourakadi 2022-1} El Bourakadi, K., Asfour, B., Sakhi, Z.,
Bennai, M., \& Ouali, T. (2022). \textit{Primordial black holes and
gravitational waves in teleparallel Gravity}. The European Physical Journal
C, 82(9), 792.

\bibitem{El Bourakadi 2021-1} El Bourakadi, K., Ferricha-Alami, M., Filali,
H., Sakhi, Z., \& Bennai, M. (2021). \textit{Gravitational waves from
preheating in Gauss--Bonnet inflation}. The European Physical Journal C,
81(12), 1144.

\bibitem{Bousder 2021} Bousder, M., El Bourakadi, K., \& Bennai, M. (2021). 
\textit{Charged 4d einstein-gauss-bonnet black hole: Vacuum solutions,
cauchy horizon, thermodynamics}. Physics of the Dark Universe, 32, 100839.

\bibitem{El Bourakadi 2022-2} El Bourakadi, K., Sakhi, Z., \& Bennai, M.
(2022). \textit{Preheating constraints in }$\alpha$\textit{%
-attractor inflation and Gravitational Waves production}. International
Journal of Modern Physics A, 37(17), 2250117.

\bibitem{El Bourakadi 2021-0} El Bourakadi, K., Bousder, M., Sakhi, Z., \&
Bennai, M. (2021). \textit{Preheating and reheating constraints in
supersymmetric braneworld inflation}. The European Physical Journal Plus,
136(8), 1-19.

\bibitem{Sakhi 2020} Sakhi, Z., El Bourakadi, K., Safsafi, A.,
Ferricha-Alami, M., Chakir, H., \& Bennai, M. (2020). \textit{Effect of
brane tension on reheating parameters in small field inflation according to
Planck-2018 data}. International Journal of Modern Physics A, 35(30),
2050191.

\bibitem{El Bourakadi 2024-1} El Bourakadi, K., Ferricha-Alami, M., Sakhi,
Z., Bennai, M., \& Chakir, H. (2024). \textit{Dark matter via baryogenesis:
Affleck--Dine mechanism in the minimal supersymmetric standard model}.
Modern Physics Letters A, 39(14), 2450060.

\bibitem{El Bourakadi 2024-2} El Bourakadi, K., Otalora, G.,
Burton-Villalobos, A., Chakir, H., Ferricha-Alami, M., \& Bennai, M. (2024). 
\textit{Insights into gravitinos abundance, cosmic strings and stochastic
gravitational wave background}. The European Physical Journal C, 84(11),
1209.

\bibitem{Hanin 2023} Hanin, A., El Bourakadi, K., Ferricha-Alami, M., Sakhi,
Z., \& Bennai, M. (2023). Reheating Mechanism from Tree Level Potential in
Standard Cosmology. International Journal of Theoretical Physics, 62(7), 143.

\bibitem{El Bourakadi 2024-3} El Bourakadi, K., Chakir, H., \& Khlopov, M.
Y. (2024). \textit{Leptogenesis effects on the gravitational waves
background: interpreting the NANOGrav measurements and JWST constraints on
primordialblack holes}. Journal of Cosmology and Astroparticle Physics,
2024(07), 018.

\bibitem{El Bourakadi 2026} El Bourakadi, K., Khlopov, M. Y., Krasnov, M.,
Chakir, H., \& Bennai, M. (2026). Tracing Inflationary Imprints Through the
Dark Ages: Implications for Early Stars and Galaxies Formation. Physics of
the Dark Universe, 102214.

\bibitem{Planck 2020} Aghanim, N., Akrami, Y., Ashdown, M., Aumont, J.,
Baccigalupi, C., Ballardini, M., ... \& Roudier, G. (2020). \textit{Planck
2018 results}. VI. Cosmological parameters.

\bibitem{Larson 2011} Larson, D., Dunkley, J., Hinshaw, G., Komatsu, E.,
Nolta, M. R., Bennett, C. L., ... \& Wright, E. L. (2011). \textit{%
Seven-year wilkinson microwave anisotropy probe (WMAP*) observations: power
spectra and WMAP-derived parameter}s. The Astrophysical Journal Supplement
Series, 192(2), 16.

\bibitem{Tegmark 1997} Tegmark, M., Silk, J., Rees, M. J., Blanchard, A.,
Abel, T., \& Palla, F. (1997). \textit{How small were the first cosmological
objects?}. The Astrophysical Journal, 474(1), 1.

\bibitem{Galli 1998} Galli, D., \& Palla, F. (1998). \textit{The chemistry
of the early Universe}. arXiv preprint astro-ph/9803315.

\bibitem{Hollenbach 1979} Hollenbach, D., \& McKee, C. F. (1979). \textit{%
Molecule formation and infrared emission in fast interstellar shocks. I
Physical processes}. Astrophysical Journal Supplement Series, vol. 41, Nov.
1979, p. 555-592., 41, 555-592.

\bibitem{Abel 2002} Abel, T., Bryan, G. L., \& Norman, M. L. (2002). \textit{%
The formation of the first star in the universe}. science, 295(5552), 93-98.

\bibitem{Bromm 2004} Bromm, V., \& Larson, R. B. (2004). \textit{The first
stars}. ARA\&A 42: 79--118. doi: 10.1146/annurev. astro. 42.053102. 134034.
arXiv preprint astro-ph/0311019.

\bibitem{Yoshida 2006} Yoshida, N., Omukai, K., Hernquist, L., \& Abel, T.
(2006). \textit{Formation of primordial stars in a }$\mathit{\Lambda }$%
\textit{CDM universe}. The Astrophysical Journal, 652(1), 6.

\bibitem{Palla 1983} Palla, F., Salpeter, E. E., \& Stahler, S. W. (1983). 
\textit{Primordial star formation-The role of molecular hydrogen}.
Astrophysical Journal, Part 1 (ISSN 0004-637X), vol. 271, Aug. 15, 1983, p.
632-641. Research supported by the Consiglio Nazionale delle Ricerche., 271,
632-641.

\bibitem{Greif 2011} Greif, T. H., Springel, V., White, S. D., Glover, S.
C., Clark, P. C., Smith, R. J., ... \& Bromm, V. (2011). \textit{Simulations
on a moving mesh: the clustered formation of population III protostars.} The
Astrophysical Journal, 737(2), 75.

\bibitem{Clark 2011} Clark, P. C., Glover, S. C., Smith, R. J., Greif, T.
H., Klessen, R. S., \& Bromm, V. (2011). \textit{The formation and
fragmentation of disks around primordial protostars}. Science, 331(6020),
1040-1042.

\bibitem{Bromm 2002} Bromm, V., Coppi, P. S., \& Larson, R. B. (2002). 
\textit{The formation of the first stars. I. The primordial star-forming
cloud}. The Astrophysical Journal, 564(1), 23.

\bibitem{Wise 2011} Wise, J. H., Turk, M. J., Norman, M. L., \& Abel, T.
(2011). \textit{The birth of a galaxy: primordial metal enrichment and
stellar populations}. The Astrophysical Journal, 745(1), 50.

\bibitem{Kreckel 2010} Kreckel, H., Bruhns, H., \v{C}\'{\i}\v{z}ek, M.,
Glover, S. C. O., Miller, K. A., Urbain, X., \& Savin, D. W. (2010).
Experimental results for H2 formation from H- and H and implications for
first star formation. Science, 329(5987), 69-71.

\bibitem{Greif 2012} Greif, T. H., Bromm, V., Clark, P. C., Glover, S. C.,
Smith, R. J., Klessen, R. S., ... \& Springel, V. (2012). \textit{Formation
and evolution of primordial protostellar systems}. Monthly Notices of the
Royal Astronomical Society, 424(1), 399-415.

\bibitem{Omukai 1998} Omukai, K., \& Nishi, R. (1998). \textit{Formation of
primordial protostars}. The Astrophysical Journal, 508(1), 141.

\bibitem{Stacy 2013} Stacy, A., \& Bromm, V. (2013). \textit{Constraining
the statistics of Population III binaries}. Monthly Notices of the Royal
Astronomical Society, 433(2), 1094-1107.

\bibitem{Stahler 1986} Stahler, S. W., Palla, F., \& Salpeter, E. E. (1986). 
\textit{Primordial stellar evolution-The protostar phase}. Astrophysical
Journal, Part 1 (ISSN 0004-637X), vol. 302, March 15, 1986, p. 590-605.
CNR-supported research., 302, 590-605.

\bibitem{Khlopov 2002} Khlopov, M. Y., \& Rubin, S. G. (2002). \textit{%
Strong primordial inhomogeneities and galaxy formation}. In Cosmological
Pattern of Microphysics in the Inflationary Universe (pp. 49-85). Dordrecht:
Springer Netherlands.

\bibitem{Khlopov 2005} Khlopov, M. Y., Rubin, S. G., \& Sakharov, A. S.
(2005). \textit{Primordial structure of massive black hole clusters}.
Astroparticle Physics, 23(2), 265-277.

\bibitem{Khlopov 2007} Khlopov, M. (2007, May). \textit{Primordial nonlinear
structures and massive black holes from early Universe}. In Journal of
Physics: Conference Series (Vol. 66, No. 1, p. 012032). IOP Publishing.

\bibitem{Barbieri 2024} Barbieri, J., \& Chapline, G. (2024). \textit{%
Predictions for a Low-mass Cutoff for the Primordial Black Hole Mass Spectrum%
}. arXiv preprint arXiv:2406.06827.

\bibitem{De Luca 2020} De Luca, V., Franciolini, G., Pani, P., \& Riotto, A.
(2020). \textit{Constraints on primordial black holes: the importance of
accretion}. Physical Review D, 102(4), 043505.

\bibitem{De Luca 2023} De Luca, V., Franciolini, G., \& Riotto, A. (2023). 
\textit{Heavy primordial black holes from strongly clustered light black
holes}. Physical Review Letters, 130(17), 171401.

\bibitem{Yuan 2024} Yuan, G. W., Lei, L., Wang, Y. Z., Wang, B., Wang, Y.
Y., Chen, C., ... \& Fan, Y. Z. (2024). \textit{Rapidly growing primordial
black holes as seeds of the massive high-redshift JWST Galaxies}. Science
China Physics, Mechanics \& Astronomy, 67(10), 109512.

\bibitem{Hoyle 1966} Hoyle, F., \& Narlikar, J. V. (1966). \textit{On the
formation of elliptical galaxies}. Proceedings of the Royal Society of
London. Series A. Mathematical and Physical Sciences, 290(1421), 177-185.

\bibitem{Carr 1983} Carr, B. J., \& Silk, J. (1983). \textit{Can graininess
in the early universe make galaxies?}. Astrophysical Journal, Part 1 (ISSN
0004-637X), vol. 268, May 1, 1983, p. 1-16., 268, 1-16.

\bibitem{Carr 1984} Carr, B. J., \& Rees, M. J. (1984). \textit{Can
pregalactic objects generate galaxies?}. Monthly Notices of the Royal
Astronomical Society, 206(4), 801-818.

\bibitem{Carr 2018} Carr, B., \& Silk, J. (2018). \textit{Primordial black
holes as generators of cosmic structures}. Monthly Notices of the Royal
Astronomical Society, 478(3), 3756-3775.

\bibitem{Sanchez 1997} Sanchez, N., \& Zichichi, A. (1997). \textit{Current
topics in astrofundamental physics}. Current topics in astrofundamental
physics/edited by N. Sanchez.

\bibitem{Green 1997} Green, A. M., \& Liddle, A. R. (1997). \textit{%
Constraints on the density perturbation spectrum from primordial black holes}%
. Physical Review D, 56(10), 6166.

\bibitem{Su 2023} Su, B. Y., Li, N., \& Feng, L. (2023). \textit{An
inflation model for massive primordial black holes to interpret the JWST
observations}. arXiv preprint arXiv:2306.05364.

\bibitem{Murgia 2019} Murgia, R., Scelfo, G., Viel, M., \& Raccanelli, A.
(2019). \textit{Lyman-}$\alpha $\textit{\ forest constraints on primordial
black holes as dark matter}. Physical review letters, 123(7), 071102.

\bibitem{Labbe 2023} Labb$\mathit{\acute{e}}$, I., van Dokkum, P., Nelson,
E., Bezanson, R., Suess, K. A., Leja, J., ... \& Wang, B. (2023). \textit{A
population of red candidate massive galaxies}$\mathit{\sim }$\textit{600 Myr
after the Big Bang}. Nature, 616(7956), 266-269.

\bibitem{Terao 2022} Terao, K., Nagao, T., Onishi, K., Matsuoka, K.,
Akiyama, M., Matsuoka, Y., \& Yamashita, T. (2022). \textit{Multiline
assessment of narrow-line regions in} $z\sim 3$ \textit{radio galaxies}. The
Astrophysical Journal, 929(1), 51.

\bibitem{Volonteri 2012} Volonteri, M., \& Bellovary, J. (2012). \textit{%
Black holes in the early Universe}. Reports on Progress in Physics, 75(12),
124901.

\bibitem{Curti 2023} Curti, M., d'Eugenio, F., Carniani, S., Maiolino, R.,
Sandles, L., Witstok, J., ... \& Wallace, I. E. (2023). \textit{The chemical
enrichment in the early Universe as probed by JWST via direct metallicity
measurements at} $z\sim 8$. Monthly Notices of the Royal Astronomical
Society, 518(1), 425-438.

\bibitem{Bouwens 2010} Bouwens, R. J., Illingworth, G. D., Gonzalez, V.,
Labbe, I., Franx, M., Conselice, C. J., ... \& Zheng, W. (2010). $z\sim 7$ 
\textit{galaxy candidates from NICMOS observations over the HDF South and
the CDF-S and HDF-N GOODS fields}. arXiv preprint arXiv:1003.1706.

\bibitem{Maiolino 2024} Maiolino, R., Scholtz, J., Witstok, J., Carniani,
S., D'Eugenio, F., de Graaff, A., ... \& Sun, F. (2024). \textit{A small and
vigorous black hole in the early Universe}. Nature, 627(8002), 59-63.

\bibitem{Fall 1980} Fall, S. M., \& Efstathiou, G. (1980). \textit{Formation
and rotation of disc galaxies with haloes}. Monthly Notices of the Royal
Astronomical Society, 193(2), 189-206.

\bibitem{Dalcanton 1997} Dalcanton, J. J., Spergel, D. N., \& Summers, F. J.
(1997). \textit{The formation of disk galaxies}. The Astrophysical Journal,
482(2), 659.

\bibitem{Mo 1998} Mo, H. J., Mao, S., \& White, S. D. (1998). \textit{The
formation of galactic discs}. Monthly Notices of the Royal Astronomical
Society, 295(2), 319-336.

\bibitem{Valentini 2020} Valentini, M., Murante, G., Borgani, S., Granato,
G. L., Monaco, P., Brighenti, F., ... \& Lapi, A. (2020). \textit{Impact of
AGN feedback on galaxies and their multiphase ISM across cosmic time}.
Monthly Notices of the Royal Astronomical Society, 491(2), 2779-2807.

\bibitem{Eddington 2013} Eddington, A. S. (2013). 45. \textit{The Internal
Constitution of the Stars}. In A Source Book in Astronomy and Astrophysics,
1900--1975 (pp. 281-290). Harvard University Press.

\bibitem{Silverstein 2008} Silverstein, E., \& Westphal, A. (2008). \textit{%
Monodromy in the CMB: gravity waves and string inflation}. Physical Review
D---Particles, Fields, Gravitation, and Cosmology, 78(10), 106003.

\bibitem{Flauger 2010} Flauger, R., McAllister, L., Pajer, E., Westphal, A.,
\& Xu, G. (2010). \textit{Oscillations in the CMB from axion monodromy
inflation}. Journal of Cosmology and Astroparticle Physics, 2010(06), 009.

\bibitem{Easther 2014} Easther, R., \& Flauger, R. (2014). Planck
constraints on monodromy inflation. Journal of Cosmology and Astroparticle
Physics, 2014(02), 037.

\bibitem{Flauger 2017} Flauger, R., McAllister, L., Silverstein, E., \&
Westphal, A. (2017). Drifting oscillations in axion monodromy. Journal of
Cosmology and Astroparticle Physics, 2017(10), 055.

\bibitem{Inomata 2017} Inomata, K., Kawasaki, M., Mukaida, K., Tada, Y., \&
Yanagida, T. T. (2017). \textit{O(10) }$M_{\odot }$\textit{\ primordial
black holes and string axion dark matter}. Physical Review D, 96(12), 123527.

\bibitem{Inomata 2021} Inomata, K. (2021). \textit{Bound on induced
gravitational waves during inflation era}. Physical Review D, 104(12),
123525.

\bibitem{Ballesteros 2018} Ballesteros, G., \& Taoso, M. (2018). \textit{%
Primordial black hole dark matter from single field inflation.} Physical
Review D, 97(2), 023501.

\bibitem{Ozsoy 2018} $\ddot{O}$zsoy, O., Parameswaran, S., Tasinato, G., \&
Zavala, I. (2018). \textit{Mechanisms for primordial black hole production
in string theory}. Journal of Cosmology and Astroparticle Physics, 2018(07),
005.

\bibitem{Gilman 2021} Gilman, D., Bovy, J., Treu, T., Nierenberg, A.,
Birrer, S., Benson, A., \& Sameie, O. (2021). Strong lensing signatures of
self-interacting dark matter in low-mass haloes. Monthly Notices of the
Royal Astronomical Society, 507(2), 2432-2447.

\bibitem{Inayoshi 2020} Inayoshi, K., Visbal, E., \& Haiman, Z. (2020). 
\textit{The assembly of the first massive black holes}. Annual Review of
Astronomy and Astrophysics, 58(1), 27-97.

\bibitem{Ellis 1993} Ellis, G. F., \& Rothman, T. (1993). \textit{Lost
horizons}. American Journal of Physics, 61(10), 883-893.

\bibitem{Weinberg 2008} Weinberg, S. (2008). \textit{Cosmology}. OUP Oxford.

\bibitem{Hu 1996} Hu, W., \& Sugiyama, N. (1996). \textit{Small-scale
cosmological perturbations: an analytic approach}. The Astrophysical
Journal, 471(2), 542.

\bibitem{Ma 1995} Ma, C. P., \& Bertschinger, E. (1995). \textit{%
Cosmological perturbation theory in the synchronous and conformal Newtonian
gauges}. arXiv preprint astro-ph/9506072.

\bibitem{Dodelson 2024} Dodelson, S., \& Schmidt, F. (2024). \textit{Modern
cosmology}. Elsevier.

\bibitem{Meszaros 1974} Meszaros, P. (1974). \textit{The behaviour of point
masses in an expanding cosmological substratum}. Astronomy and Astrophysics,
vol. 37, no. 2, Dec. 1974, p. 225-228., 37, 225-228.

\bibitem{Blas 2011} Blas, D., Lesgourgues, J., \& Tram, T. (2011). \textit{%
The cosmic linear anisotropy solving system (CLASS). Part II: Approximation
schemes}. Journal of Cosmology and Astroparticle Physics, 2011(07), 034.

\bibitem{Jeans 1902} Jeans, J. H. (1902). I. \textit{The stability of a
spherical nebula}. Philosophical Transactions of the Royal Society of
London. Series A, Containing Papers of a Mathematical or Physical Character,
199(312-320), 1-53.

\bibitem{Schaye 2008} Schaye, J., \& Dalla Vecchia, C. (2008). \textit{On
the relation between the Schmidt and Kennicutt--Schmidt star formation laws
and its implications for numerical simulations}. Monthly Notices of the
Royal Astronomical Society, 383(3), 1210-1222.

\bibitem{Tseliakhovich 2010a} Tseliakhovich, D., \& Hirata, C. (2010). 
\textit{Relative velocity of dark matter and baryonic fluids and the
formation of the first structures}. Physical Review D---Particles, Fields,
Gravitation, and Cosmology, 82(8), 083520.

\bibitem{Tseliakhovich 2010b} Tseliakhovich, D., Hirata, C., \& Slosar, A.
(2010). \textit{Non-Gaussianity and large-scale structure in a two-field
inflationary model.} Physical Review D---Particles, Fields, Gravitation, and
Cosmology, 82(4), 043531.

\bibitem{Hirano 2014} Hirano, S., Hosokawa, T., Yoshida, N., Umeda, H.,
Omukai, K., Chiaki, G., \& Yorke, H. W. (2014). \textit{One hundred first
stars: protostellar evolution and the final masses}. The Astrophysical
Journal, 781(2), 60.

\bibitem{Stahl 2025} Stahl, C., Werth, D., \& Poulin, V. (2025). \textit{%
Primordial Sharp Features through the Nonlinear Regime of Structure Formation%
}. arXiv preprint arXiv:2502.02571.

\bibitem{Liu 2021} Liu, Y., Wang, R., Momjian, E., Ba\~{n}ados, E., Zeimann,
G., Willott, C. J., ... \& Li, J. (2021). \textit{Constraining the Quasar
Radio-loud Fraction at }$\left( \mathit{z\sim 6}\right) $\textit{\ with Deep
Radio Observations}. The Astrophysical Journal, 908(2), 124.

\bibitem{Lesgourgues 2014} Lesgourgues, J., \& Tram, T. (2014). \textit{Fast
and accurate CMB computations in non-flat FLRW universes}. Journal of
Cosmology and Astroparticle Physics, 2014(09), 032.

\bibitem{Munoz 2020} Mu$\tilde{n}$oz, J. B., Dvorkin, C., \& Cyr-Racine, F.
Y. (2020). \textit{Probing the small-scale matter power spectrum with
large-scale 21-cm data}. Physical Review D, 101(6), 063526.

\bibitem{Chen 2016} Chen, X., Meerburg, P. D., \& M\"{u}nchmeyer, M. (2016).
The future of primordial features with 21 cm tomography. Journal of
Cosmology and Astroparticle Physics, 2016(09), 023.

\bibitem{Bennett 1996} Bennett, C. L., Banday, A. J., G\'{o}rski, K. M.,
Hinshaw, G., Jackson, P., Keegstra, P., ... \& Wright, E. L. (1996). \textit{%
Four-year COBE* DMR cosmic microwave background observations: maps and basic
results}. The Astrophysical Journal, 464(1), L1.

\bibitem{Peebles 1989} Peebles, P. J. E. (1989). \textit{Tracing galaxy
orbits back in time}. Astrophysical Journal, Part 2-Letters (ISSN
0004-637X), vol. 344, Sept. 15, 1989, p. L53-L56. Research supported by
NSF., 344, L53-L56.

\bibitem{Bullock 2017} Bullock, J. S., \& Boylan-Kolchin, M. (2017). \textit{%
Small-scale challenges to the }$\mathit{\Lambda }$\textit{CDM paradigm}.
Annual Review of Astronomy and Astrophysics, 55(1), 343-387.

\bibitem{Cen 1992} Cen, R., \& Ostriker, J. (1992). \textit{A hydrodynamic
treatment of the cold dark matter cosmological scenario}. Astrophysical
Journal, Part 1 (ISSN 0004-637X), vol. 393, no. 1, July 1, 1992, p. 22-41.,
393, 22-41.

\bibitem{Zel'dovich 1970} Zel'dovich, Y. B. (1970). \textit{The Hypothesis
of Cosmological Magnetic Inhomogeneity.} Soviet Astronomy, Vol. 13, p. 608,
13, 608.

\bibitem{Hahn 2016} Hahn, O., \& Angulo, R. E. (2016). \textit{An adaptively
refined phase--space element method for cosmological simulations and
collisionless dynamics}. Monthly Notices of the Royal Astronomical Society,
455(1), 1115-1133.

\bibitem{Ryu 1993} Ryu, D., Ostriker, J. P., Kang, H., \& Cen, R. (1993). 
\textit{A cosmological hydrodynamic code based on the total variation
diminishing scheme.} Astrophysical Journal, Part 1 (ISSN 0004-637X), vol.
414, no. 1, p. 1-19., 414, 1-19.

\bibitem{Kolb 1990} Kolb, E. W., Salopek, D. S., \& Turner, M. S. (1990). 
\textit{Origin of density fluctuations in extended inflation}. Physical
Review D, 42(12), 3925.

\bibitem{Peebles 2020} Peebles, P. J. E. (2020). \textit{The large-scale
structure of the universe} (Vol. 96). Princeton university press.

\bibitem{Press 1974p} Press, W. H., \& Schechter, P. (1974). \textit{Remark
on the statistical significance of flares in Poisson count data}.
Astrophysical Journal, Vol. 193, pp. 437-442 (1974), 193, 437-442.

\bibitem{Carroll 1992} Carroll, S. M., Press, W. H., \& Turner, E. L.
(1992). \textit{The cosmological constant}. In: Annual review of astronomy
and astrophysics. Vol. 30 (A93-25826 09-90), p. 499-542., 30, 499-542.

\bibitem{Jenkins 2001} Jenkins, A., Frenk, C. S., White, S. D., Colberg, J.
M., Cole, S., Evrard, A. E., ... \& Yoshida, N. (2001). \textit{The mass
function of dark matter haloes}. Monthly Notices of the Royal Astronomical
Society, 321(2), 372-384.

\bibitem{Ellis 1999} Ellis, G. F., \& Van Elst, H. (1999). \textit{%
Cosmological models: Cargese lectures 1998}. Theoretical and Observational
Cosmology, 1-116.

\bibitem{Yoshida 2003} Yoshida, N., Abel, T., Hernquist, L., \& Sugiyama, N.
(2003). \textit{Simulations of early structure formation: Primordial gas
clouds}. The Astrophysical Journal, 592(2), 645.

\bibitem{Wise 2008} Wise, J. H., Turk, M. J., \& Abel, T. (2008). \textit{%
Resolving the formation of protogalaxies. II. Central gravitational collapse}%
. The Astrophysical Journal, 682(2), 745.

\bibitem{Gould 1963} Gould, R. J., \& Salpeter, E. E. (1963). \textit{The
Interstellar Abundance of the Hydrogen Molecule. I. Basic Processes}.
Astrophysical Journal, vol. 138, p. 393, 138, 393.

\bibitem{Gould I 1963} Gould, R. J., Gold, T., \& Salpeter, E. E. (1963). 
\textit{The Interstellar Abundance of the Hydrogen Molecule. II. Galactic
Abundance and Distribution}. Astrophysical Journal, vol. 138, p. 408, 138,
408.

\bibitem{Tielens 1985} Tielens, A. G. G. M., \& Hollenbach, D. (1985). 
\textit{Photodissociation regions. I-Basic model. II-A model for the Orion
photodissociation region}. Astrophysical Journal, Part 1 (ISSN 0004-637X),
vol. 291, April 15, 1985, p. 722-754. NASA-supported research., 291, 722-754.

\bibitem{Khlopov 2010} Khlopov, M. Y. (2010). \textit{Primordial black holes}%
. Research in Astronomy and Astrophysics, 10(6), 495.

\bibitem{Brockamp 2016} Brockamp, M., Baumgardt, H., Britzen, S., \& Zensus,
A. (2016). \textit{Unveiling Gargantua: A new search strategy for the most
massive central cluster black holes}. Astronomy \& Astrophysics, 585, A153.

\bibitem{Inoue 2017} Inoue, Y., \& Kusenko, A. (2017). \textit{New X-ray
bound on density of primordial black holes.} Journal of Cosmology and
Astroparticle Physics, 2017(10), 034.

\bibitem{Carr 1999} Carr, B. J., \& Sakellariadou, M. (1999). \textit{%
Dynamical constraints on dark matter in compact objects.} The Astrophysical
Journal, 516(1), 195.

\bibitem{Boylan 2025} Boylan-Kolchin, M. (2025). \textit{Accelerated by dark
matter: a high-redshift pathway to efficient galaxy-scale star formation}.
Monthly Notices of the Royal Astronomical Society, 538(4), 3210-3218.

\bibitem{Fan 2023} Fan, X., Ba\~{n}ados, E., \& Simcoe, R. A. (2023). 
\textit{Quasars and the intergalactic medium at cosmic dawn}. Annual Review
of Astronomy and Astrophysics, 61(1), 373-426.

\bibitem{Pensabene 2021} Pensabene, A., Decarli, R. O. B. E. R. T. O., Ba%
\~{n}ados, E., Venemans, B., Walter, F., Bertoldi, F., ... \& Yang, Y.
(2021). \textit{ALMA multiline survey of the ISM in two quasar
host--companion galaxy pairs at} $\mathit{z>6}$. Astronomy \& Astrophysics,
652, A66.

\bibitem{Walter 2022} Walter, F., Neeleman, M., Decarli, R., Venemans, B.,
Meyer, R., Weiss, A., ... \& Thompson, T. A. (2022). \textit{Alma 200 pc
imaging of a} $z\sim 7$ \textit{quasar reveals a compact, disk-like host
galaxy}. The Astrophysical Journal, 927(1), 21.

\bibitem{Navarro 1991} Navarro, J. F., \& Benz, W. (1991). \textit{Dynamics
of cooling gas in galactic dark halos}. Astrophysical Journal, Part 1 (ISSN
0004-637X), vol. 380, Oct. 20, 1991, p. 320-329. Research supported by SNSF
and CONICET., 380, 320-329.

\bibitem{Evrard 1994} Evrard, A. E., Summers, F. J., \& Davis, M. (1994). 
\textit{2-Fluid Simulations of Galaxy Formation}.

\bibitem{Navarro 1995} Navarro, J. F., Frenk, C. S., \& White, S. D. (1995). 
\textit{Simulations of X-ray clusters}. Monthly Notices of the Royal
Astronomical Society, 275(3), 720-740.

\bibitem{Tissera 1997} Tissera, P. B., Lambas, D. G., \& Abadi, M. G.
(1997). \textit{Analysis of galaxy formation with hydrodynamics}. Monthly
Notices of the Royal Astronomical Society, 286(2), 384-392.

\bibitem{Eke 2000} Eke, V., Efstathiou, G., \& Wright, L. (2000). \textit{%
The cosmological dependence of galactic specific angular momenta}. Monthly
Notices of the Royal Astronomical Society, 315(2), L18-L22.

\bibitem{Binney 2001} Binney, J. J., \& Evans, N. W. (2001). \textit{Cuspy
dark matter haloes and the Galaxy}. Monthly Notices of the Royal
Astronomical Society, 327(2), L27-L31.

\bibitem{Efstathiou 2000} Efstathiou, G. (2000). \textit{A model of
supernova feedback in galaxy formation}. Monthly Notices of the Royal
Astronomical Society, 317(3), 697-719.

\bibitem{Navarro 2000} Navarro, J. F., \& Steinmetz, M. (2000). \textit{Dark
Halo and Disk Galaxy Scaling Laws in Hierarchical Universes}. The
Astrophysical Journal, 538(2), 477.

\bibitem{PPNP} Khlopov, M. (2021) \textit{\ What comes after the Standard
model?} Progress in Particle and Nuclear Physics, 116, 103824.

\bibitem{Guo2024} Guo, S.-Y., Khlopov, M., Liu, X., Wu, L., Wu, Y. \& Zhu,
B. (2024) \textit{Footprints of Axion-Like Particle in Pulsar Timing Array
Data and JWST Observations}. SCIENCE CHINA Physics, Mechanics $\&$
Astronomy, 67, 111011
\end{thebibliography}
\end{document}